\documentclass[11pt]{article}
\usepackage{amsmath,amssymb}

\newcommand{\reseteqnum}{\setcounter{equation}{0}}

\newcommand{\fs}[1]{{#1 \hskip -7.5pt /}}
\newcommand{\fsu}[1]{{#1 \hskip -7.5pt /}}
\newcommand{\larrow}{\mathop{\longrightarrow}}

\newcommand{\ce}{\mathop{=}}

\title{
\hfill
\parbox{3cm}{\normalsize KUNS-1574 DPNU-99-16\\
{\tt  hep-lat/9905003}}\\
\vspace{0.5cm}
A lattice implementation of the $\eta$-invariant and\\
effective action for chiral fermions on the lattice
\author{
Tatsumi Aoyama\thanks{e-mail address:
aoyama@gauge.scphys.kyoto-u.ac.jp} 
\\
{\normalsize\em Department of Physics, Kyoto University }\\
{\normalsize\em Kyoto 606-8502, Japan}
\\
\\
and
\\
\\
Yoshio Kikukawa\thanks{e-mail address:
kikukawa@eken.phys.nagoya-u.ac.jp} 
\\
{\normalsize\em Department of Physics, Nagoya University }\\
{\normalsize\em Nagoya 464-8602, Japan}
\\
\date{\normalsize May, 1999}
}
}

\begin{document}
\maketitle

\begin{abstract}
We consider a lattice implementation of the $\eta$-invariant, 
using the complex phase of the determinant of the simplified 
domain-wall fermion, which couples to an interpolating 
five-dimensional gauge field. We clarify the relation to the 
effective action for chiral Ginsparg-Wilson fermions. The 
integrability, which holds true for anomaly-free theories in the 
classical continuum limit, is not assured on the lattice with a 
finite spacing. A lattice expression for the five-dimensional 
Chern-Simons term is obtained. 
\end{abstract}

\newpage
\section{Introduction}
\label{sec:intro}
\reseteqnum

It has become clear recently that the gauge interaction of the Weyl 
fermions can be described in the framework of lattice gauge theory.
The clue to this development is the construction of gauge covariant 
and local Dirac operators \cite{overlap-D,
fixed-point-D,locality-of-overlap-D} which solve the Ginsparg-Wilson relation 
\cite{ginsparg-wilson-rel}. The Ginsparg-Wilson relation implies the 
exact chiral symmetry for the Dirac fermion
\cite{exact-chiral-symmetry} and suggests an asymmetric and
gauge-field-dependent chiral projection to the Weyl degrees of freedom 
\cite{neidermayer-lat98,ginsparg-wilson-relation-and-overlap}. 
The functional measure for the Weyl fermion field is defined based
on the chiral projection. It leads to a mathematically reasonable
definition of the chiral determinant, which generally has the
structure as an overlap of two vacua \cite{overlap}. It has been 
shown by L\"uscher in \cite{topology-and-axial-anomaly,
abelian-chiral-gauge-theory} that for anomaly-free abelian chiral 
gauge theories, the functional measure for 
the Weyl fermion fields can be constructed so that the gauge
invariance is maintained exactly on the lattice.

On the other hand, in the continuum theory, 
Alvarez-Gaum\'e et al. \cite{eta-invariant} and 
Ball and Osborn \cite{ball-osbor} 
have shown that the imaginary part of the
effective action for chiral fermions can be given by the
$\eta$-invariant \cite{atiyah-padoti-singer}.
It is defined as the spectrum asymmetry of 
the five-dimensional massless Dirac 
operator coupled to an interpolating five-dimensional gauge field. 

In this paper, we will show that a lattice implementation of the 
$\eta$-invariant is possible so that the lattice $\eta$-invariant
gives the imaginary part of the effective action for the chiral 
Ginsparg-Wilson fermion defined by Neuberger's Dirac operator.
We define the $\eta$-invariant on the lattice using the complex phase
of the determinant of the (simplified) domain-wall fermion 
\cite{domain-wall-fermion,boundary-fermion}, which couples to an 
interpolating five-dimensional gauge field. Our formulation then can be 
regarded as a lattice realization of the argument 
given by Kaplan and Shmaltz in the continuum theory
\cite{domain-wall-ferimon-and-eta-invariant}, using the simplified 
formulation of the domain-wall fermion by Shamir \cite{boundary-fermion}. 

Our lattice implementation of the $\eta$-invariant can be shown to have 
a direct relation to the imaginary part of the effective action for
the chiral Ginsparg-Wilson fermions which is defined by
Neuberger's Dirac 
operator \cite{overlap,abelian-chiral-gauge-theory,suzuki}. 
This implementation is applicable to non-abelian chiral gauge theories. 
But the integrability, which holds true 
for anomaly-free theories in the classical continuum limit, 
is not assured 
on the lattice with a finite spacing. 
This issue of the integrability for anomaly free chiral gauge theories 
is discussed. A lattice expression for the five-dimensional
Chern-Simons term is obtained. \footnote{When completing this work 
and preparing this article, we noticed that a paper by L\"uscher 
\cite{non-abelian-gauge-anomaly} appeared. 
In \cite{non-abelian-gauge-anomaly}, a formula of the effective action 
for the chiral Ginsparg-Wilson fermion which couples to non-abelian 
gauge fields is derived and its relation to the $\eta$-invariant is 
suggested. }

This paper is organized as follows.
The section~\ref{sec:reviews}
is devoted to reviews of effective action for chiral fermions
in the continuum theory and on the lattice:
In section~\ref{sec:eta-invariant-in-continuum},
we first review the relation bewteen
the effective action for chiral fermions 
and the $\eta$-invariant in the continuum theory. 
In section~\ref{sec:effective-action}, 
the effective action for the chiral Ginsparg-Wilson fermion 
defined through Neuberger's lattice Dirac operator is reviewed. 
In section~\ref{sec:domain-wall-fermion}, 
we discuss the relation between 
Neuberger's Dirac operator and the domain-wall fermion 
for vector-like theories.
In section~\ref{sec:implementation}, we describe our implementation
of the $\eta$-invariant on the lattice, using chiral 
domain-wall fermion.
In section~\ref{sec:variation}, we examine the variation of the
lattice $\eta$-invariant with respect to the gauge field. 
In section~\ref{sec:relation-to-effective-action}, 
we clarify the relation of the 
lattice $\eta$-invariant to the effective action for chiral
Ginsparg-Wilson fermions
defined through Neuberger's lattice Dirac operator. 
The integrability for anomaly free chiral gauge theories, which is 
not assured a priori on the lattice, is discussed.
In section~\ref{sec:summary}, we summarize our result and give some 
discussions.

In the following discussions, 
various fermion theories are considered 
in the continuum limit, and on four- and five- dimensional lattices.
The four-dimensional space-time coordinates are denoted by 
$x_\mu (\mu=1,2,3,4)$ and the fifth-dimensional coordinate is
denoted by $t$.
We denote the lattice spacing of the four direction $\mu=1,2,3,4$ 
with $a$ and the lattice spacing of the fifth direction with $a_5$.
Then the lattice indices are given as follows:
\begin{eqnarray}
x_\mu &=& n_\mu \, a , \quad \, \, n_\mu 
\in \mathbb{Z}  
\quad (\mu=1,2,3,4),
\\
t &=& n_5 \, a_5 , \quad n_5 
\in \mathbb{Z} .
\end{eqnarray}

The gauge-covariant difference operators are defined with
link variables as
\begin{eqnarray}
\nabla_\mu \phi(x,t) 
&=& \frac{1}{a}\left( U_\mu(x,t) \phi(x+\hat\mu a,t) - \phi(x,t) \right),
\quad (\mu=1,2,3,4) \\
\nabla_5 \phi(x,t) 
&=& \frac{1}{a_5}\left( U_5(x,t) \phi(x,t+a_5) - \phi(x,t) \right),
\end{eqnarray}
where the unit vector in the direction $\mu$ is denoted by $\hat \mu$.

\section{Effective action for chiral fermions \\ 
in the continuum theory and on the lattice}
\label{sec:reviews}
\reseteqnum

\subsection{The $\eta$-invariant in the continuum theory}
\label{sec:eta-invariant-in-continuum}

In the continuum theory, the $\eta$-invariant
\cite{atiyah-padoti-singer}
is defined as the spectrum asymmetry of the hermitian 
five-dimensional Dirac operator.
It can be defined through the complex phase of the 
determinant of a five-dimensional 
massless Dirac fermion in the Pauli-Villars regularization 
\cite{eta-invariant,anomaly-in-odd-dimensions}.  
\begin{equation}
  \pi \eta = - 
{\rm Im} 
\log \det \left[
   \left(H - i M \right) \left / \right.
   \left(H + i M \right) \right],
\qquad H \equiv i\sum_{M=1}^5 \gamma_M D_M .
\end{equation}
In this formula, it is assumed that the five-dimensional Dirac 
operator couples to gauge fields $A_M(x,t)$ with the following property:
for $t=-\infty$ to $-\Delta$, $A_M(x,t) = A^0_M(x)$;
for $t=-\Delta$ to $+\Delta$, $A_M(x,t)$ smoothly interpoltes
between $A^0_M(x)$ and $A^1_M(x)$ and from 
$t=+\Delta$ to $+\infty$, $A_M(x,t) = A^1_M(x)$.
Both $A^0_M(x)$ and $A^1_M(x)$ are assumed to be 
perturbative configurations and $A_M(x,t)$ can be chosen so that
the five-dimensional Dirac operator does not have zero modes.

In \cite{eta-invariant}, 
the variation of the $\eta$-invariant with respect to gauge field
has been examined,
by introducing one more parameter $u$ 
which parametrizes the gauge configuration for $t \ge + \Delta$.
The result can be written as the sum of the four-dimensional surface
contribution and the five-dimensional bulk contribution:
\begin{eqnarray}
\pi \frac{d}{du} \eta
&=& {\rm Im} \,
{\rm Tr}_x \, P_L\, \frac{d}{du} \fsu D \, \frac{1}{\fs D} \nonumber\\
&-& 
\lim_{T \rightarrow \infty}  
\int d^4 x \int^{T}_{-T} dt \, 
\frac{1}{32\pi^2} \, \epsilon_{\mu MNKL} 
{\rm Tr}\left\{ \frac{d}{du} A_\mu \, F_{MN} F_{KL}\right\}(x,t;u) .
\nonumber\\
\end{eqnarray}
The first term has the form of the gauge current induced 
by chiral fermions which is regularized gauge-covariantly.
The second five-dimensional term can be written as the variation of 
the Chern-Simons term \cite{eta-invariant}, up to the local current 
of Bardeen and Zumino \cite{bardeen-zumino}:
\begin{equation}
-\frac{d}{du} \left\{ 2\pi Q_5\left[A_\mu(x,t;u)\right] \right\}
+ \int d^4 x \, 
 {\rm tr} \left\{ \frac{d}{du}A_\mu(x;u) \, X_\mu(x;u) \right\},   
\end{equation}
where 
\begin{eqnarray}
&& 2\pi Q_5\left[A_\mu\right] \nonumber\\
&& \quad
=\lim_{T \rightarrow \infty} 
\int \int^T_{-T} d^4 x \, dt \, 
\int_0^1 d \sigma \,
\frac{1}{ 32\pi^2} \epsilon_{\mu MNKL} 
{\rm Tr} \left\{ A_\mu F^\sigma_{MN} F^\sigma_{KL}\right\}, \\
&& \nonumber\\
&& \quad 
F^\sigma_{MN} = \sigma \left( \partial_M A_N - \partial_N A_M \right)
+ \sigma^2 i \left[ A_M, A_N \right] ,
\end{eqnarray}
and 
\begin{equation}
X_\mu = 
\frac{1}{48\pi^2} \,\epsilon_{\mu \nu \lambda \rho}
\left( A_\nu F_{\lambda\rho}+F_{\nu \lambda}A_\rho
      - A_\nu A_\lambda A_\rho \right) .
\end{equation}
The current of Bardeen and Zumino, denoted by $X_\mu$ here, 
plays the role to translate the covariant gauge current, which is 
induced from the surface term, 
to the consistent current \cite{eta-invariant}.

Then integrating the expression with respect to $u$,
\begin{eqnarray}
\pi \frac{d}{du} {\eta}
&=& {\rm Im} \,
\int d^4 x \, {\rm Tr} \, \, 
i \frac{d}{du} A_\mu \,\left\{ 
{\rm tr} \left(\frac{1}{2} \gamma_\mu \gamma_5 \, \frac{1}{\fsu D} \right)
+ X_\mu \right\}  \left[A_\mu(x;u)\right] 
\nonumber\\
&-& 
\frac{d}{du} \left\{ 2\pi Q_5\left[A_\mu(x,t)\right] \right\} ,
\nonumber\\
\end{eqnarray}
we obtain
\begin{equation}
\label{eq:eta-invariant-and-effective-action-continuum}
{\rm Im} \, \Gamma_{\rm eff}\left[A^1_\mu\right]
 -{\rm Im} \, \Gamma_{\rm eff}\left[A^0_\mu\right]
= \pi {\eta} 
+ 2\pi Q_5\left[A_\mu(x,t)\right] .
\end{equation}
We note here the role of the Chern-Simons term. 
1) First of all, 
the Chern-Simons term compensates the dependence 
of the $\eta$-invariant on the path 
of the interpolation and make it integrable so that
it can give the effective action of chiral fermions which depends only
the values of gauge fields at the boundaries.
2) The Chern-Simons term reproduces the non-abelian gauge 
anomaly of the effective action, 
while the $\eta$-invariant is gauge invariant.
If $A^1_\mu(x)$ is obtained from $A^0_\mu(x)$  by a certain
gauge tranformation,
\begin{equation}
A^1_\mu(x) = g(x) A^0_\mu(x) g(x)^{-1} - i g(x) \partial_\mu g(x)^{-1} ,
\end{equation}
we may consider an interpolation of the gauge transformation
function, $g(x,t)$, such that 
$ g(x,t=-\infty)=1 $ and $ g(x,t=\infty)=g(x)$ 
and the region of the interpolation is within $t\in[-\Delta,\Delta]$.
Then we obtain
\begin{eqnarray}
\label{eq:wess-zumino-term-continuum}
&&  {\rm Im} \, {\Gamma}_{\rm eff} 
\left[{\scriptstyle 
g(x) A^0_\mu(x) g(x)^{-1} - i g(x) \partial_\mu g(x)^{-1} }
\right]
-{\rm Im} \, {\Gamma}_{\rm eff} 
\left[{\scriptstyle
A^0_\mu(x)  }
\right]
\nonumber\\
&& \qquad \quad
= 
2\pi {Q}_5 
\left[{\scriptstyle
g(x,t) A^0_\mu(x) g(x,t)^{-1} - i g(x,t) \partial_\mu g(x,t)^{-1};
 -i g(x,t) \partial_5 g(x,t)^{-1}  }
\right] . \nonumber\\
\end{eqnarray}
The r.h.s. is nothing but the Wess-Zumino action.
3) When the non-abelian gauge anomaly is canceled by the condition
\begin{equation}
\sum_R {\rm Tr}_R \left(T^a \left\{ T^b , T^c \right\}\right) = 0 ,
\end{equation}
the Chern-Simons term vanishes. The $\eta$-invariant becomes
integrable and identical to the imaginary part of the 
gauge invariant effective action for chiral fermions.
\begin{equation}
{\rm Im} \, \Gamma_{\rm eff}\left[A_\mu\right]
 -{\rm Im} \, \Gamma_{\rm eff}\left[A^0_\mu\right]
= \pi {\eta}  . 
\end{equation}
Thus the imaginary part of the effective action of chiral fermions
can be expressed through the $\eta$-invariant.

Since $X_\mu(x;u)$ is orthogonal to $A_\mu(x;u)$, it does not
contribute in the integration of $u$ if we adopt the linear 
interpolation as $A_\mu(x;u)= u A_\mu(x)$.
In this case, the imaginary part of the effective action is entirely
given by the integration of the covariant gauge current induced from 
the surface term \cite{mitra}.
\begin{equation}
\label{eq:integral-for-effective-action}
{\rm Im} \, \Gamma_{\rm eff}\left[A_\mu\right]
 -{\rm Im} \, \Gamma_{\rm eff}\left[A^0_\mu\right]
= \int^1_0 du \,   
{\rm Im} \,
{\rm Tr}_x \, P_L\, \frac{d}{du} \fsu D \, \frac{1}{\fsu D} .
\end{equation}
The integration of the bulk term gives directly the Chern-Simons term.
\begin{eqnarray}
&& 2\pi Q_5\left[A_\mu(x,t;u)\right]  \nonumber\\
&& = \int^1_0 du \,   
\lim_{T \rightarrow \infty}  
\int \int^{T}_{-T} d^4 x dt \, 
\frac{1}{32\pi^2} \, \epsilon_{\mu MNKL} 
{\rm Tr}\left\{ \frac{d}{du} A_\mu \, F_{MN} F_{KL}\right\}(x,t;u).
\nonumber\\
\end{eqnarray}

\subsection{Effective action for chiral Ginsparg-Wilson fermion \\
defined by Neuberger's lattice Dirac operator}
\label{sec:effective-action}

\subsubsection{Neuberger's lattice Dirac operator}

Neuberger's lattice Dirac operator \cite{overlap-D}, 
which is derived from the overlap formalism \cite{overlap} and 
satisfies the Ginsparg-Wilson relation \cite{ginsparg-wilson-rel}, 
is given by the following formula:
\begin{equation}
\label{eq:overlap-dirac-operator}
D
= \frac{1}{2a} \left( 1 + X \frac{1}{\sqrt{X^\dagger X}} \right)
= \frac{1}{2a} \left( 1 + \gamma_5 \frac{H}{\sqrt{H^2}} \right),
\end{equation}
where $X$ is the Wilson-Dirac operator with a negative mass
and $H$ is the hermitian operator,
\begin{equation}
  X = D_{\rm w} - \frac{m_0}{a} , 
\qquad H = \gamma_5 \left( D_{w} - \frac{m_0}{a} \right),
\qquad (0 < m_0 < 2) ,
\end{equation}
\begin{equation}
D_{\rm w}=\sum_{\mu=1}^4 \left\{ 
\gamma_\mu \frac{1}{2}\left(\nabla_\mu - \nabla_\mu^\dagger \right) 
+\frac{a}{2} \nabla_\mu \nabla_\mu^\dagger\right\}.
\end{equation}
This Dirac operator satisfies the Ginsparg-Wilson 
relation \cite{ginsparg-wilson-rel}.
\begin{equation}
\label{eq:ginsparg-wilson-relation}
\gamma_5 D +   D \gamma_5  = 2 a D \gamma_5 D .
\end{equation}

Locality properties of Neuberger's lattice Dirac operator has been 
examined by Hern\'andes, Jansen and L\"uscher \cite{locality-of-overlap-D}.
For a certain class of gauge fields with small lattice field strength,
exponential bounds have been proved rigorously on the kernels of 
the Dirac operator and its differentiations with respect to the gauge field. 
Namely, if the field strength of the gauge field bounded uniformly
as follows, 
\begin{equation}
\label{eq:smooth-gauge-field}
\left\Arrowvert  1-U_{\mu\nu}(x) \right\Arrowvert 
< \epsilon, 
\qquad 
\epsilon < \frac{1}{30}\left\{1-|1-m_0|^2\right\},
\end{equation}
then the Wilson-Dirac operator square is bounded below by a positive
constant as
\begin{equation}
\left\|  a^2 \left(D_{\rm w} - \frac{m_0}{a}\right)^\dagger 
        \left(D_{\rm w} - \frac{m_0}{a}\right) \right\|
\, \ge \,  \left\{ (1- 30 \epsilon)^{\frac{1}{2}}-|1-m_0| \right\}^2 .
\end{equation}
Given the positive lower and upper bounds for $a^2 X^\dagger X$, it
follows that the kernel of Neuberger's Dirac operator is exponentially 
bounded as 
\begin{equation}
a^4 \parallel D(x,y) \parallel 
= C \exp \left\{ -\frac{\theta}{2a} |x-y| \right\},
\end{equation}
where $\theta$ is a certain constant which is determined from the 
lower and upper bounds for $a^2 X^\dagger X$.

\subsubsection{Effective action for chiral Ginsparg-Wilson fermion}

The Ginsparg-Wilson relation Eq.~(\ref{eq:ginsparg-wilson-relation})
implies the exact symmetry of the fermion action.
For the Dirac fermion described by the lattice Dirac operator 
which satisfies the Ginsparg-Wilson relation
\begin{equation}
  S_D = a^4 \sum_x \bar \psi(x) D \psi(x) ,
\end{equation}
chiral transformation can be defined as follows:
\begin{equation}
  \delta \psi(x) = \gamma_5 \left( 1-2 a D \right) \psi(x), 
\quad \delta \bar \psi(x) = \bar \psi(x) \gamma_5.
\end{equation}
Then it is straightforward to see that 
the action is invariant under this transformation.

From this property, Weyl fermion can be introduced as the eigenstate
of the generators of the chiral transformation
\begin{equation}
  \hat \gamma_5 = \gamma_5 \left( 1-2 a D \right),
\qquad 
\gamma_5. 
\end{equation}
Namely, the right-handed Weyl fermion is defined through the constraint 
given as follows \cite{neidermayer-lat98,abelian-chiral-gauge-theory}:
\begin{equation}
  \hat P_R \psi_R(x) = \psi_R(x), \quad 
  \bar \psi_R(x) P_L = \bar \psi_R(x), 
\end{equation}
where $\hat P_R$ is the chiral projector for the fermion field 
$\psi(x)$ defined as 
\begin{equation}
  P_R =\left( \frac{1+\hat \gamma_5}{2} \right), 
\qquad
  P_L =\left( \frac{1- \gamma_5}{2} \right) .
\end{equation}
The action of the Weyl fermion is given by
\begin{equation}
  S_W = a^4 \sum_x \bar \psi_R (x) D \psi_R (x) .
\end{equation}

The functional integral measure for the Weyl fermion can be defined
by introducing the chiral basis,
\begin{equation}
\hat P_R v_j(x) = v_j(x),  
\qquad \quad
\left( v_k,v_j \right)=\delta_{kj},
\end{equation}
where $\left( v_k,v_j \right)=a^4 \sum_x v_k(x)^\dagger v_j(x)$. 
Using this chiral basis, the Weyl fermion can be expanded
with the coefficients $c_j$ which generates a Grassmann algebra.
Then the functional measure for the right-handed field is given by 
\begin{equation}
 D\left[\psi \right]= \prod_j d c_j 
\qquad   \psi_R(x) = \sum_j v_j(x) c_j .
\end{equation}
The functional measure for the anti-fermion field is 
defined similarly with the basis
\begin{equation}
\bar v_k(x) P_L = \bar v_k(x), 
\qquad \quad
\left( \bar v_k,\bar v_j \right)=\delta_{kj},
\end{equation}
as follows:
\begin{equation}
 D\left[\bar \psi \right]= \prod_k d \bar c_k
\qquad   \bar \psi_R(x) = \sum_k \bar c_k \bar v_k(x) .
\end{equation}
Then the partition function of the Weyl fermion is given as
\begin{equation}
  Z_F = \int D\left[\psi\right] D\left[\bar \psi\right]
        e^{- S_W\left[\psi_R,\bar \psi_R\right] }
      = \det M_{kj} ,
\end{equation}
where 
\begin{equation}
  M_{kj}= \bar v_k D v_j .
\end{equation}

In the case using Neuberger's Dirac operator, the right-handed 
projector is noting but the projector to negative energy states 
of the Wilson-Dirac hamiltonian $H$,
\begin{equation}
  \hat P_R = \frac{1- \frac{H}{\sqrt{H^2}}}{2}. 
\end{equation}
Then the basis $v_j(x)$ can be chosen as the normalized 
complete set of the eigenvectors of negative energy states. 
The basis $\bar v_k(x)$ may be regarded as the complete set of the 
negative-energy eigenvectors of the hamiltonian $\gamma_5$.  
Then the chiral determinant may be written in the form of the overlap 
of two vacua \cite{overlap}:
\begin{equation}
\det M_{kj} = \det \left( \bar v_k v_j \right) .
\end{equation}

It is useful for later discussions to consider the variation of 
the effective action with respect to the gauge fields 
\cite{geometrical-aspect,
abelian-chiral-gauge-theory,non-abelian-gauge-anomaly}.
Following \cite{abelian-chiral-gauge-theory,non-abelian-gauge-anomaly},
we write the variation of the link variables as
\begin{equation}
\delta_\zeta U_\mu(x)= a \zeta_\mu(x) U_\mu(x) , \qquad
\zeta_\mu(x) = i T^a \zeta^a_\mu(x).
\end{equation}
Then the variation of the effective action is evaluated as
\begin{equation}
\label{eq:effective-action-variation}
\delta_\zeta  \ln \det M_{kj} = 
{\rm Tr} \delta_\zeta D  \hat P_R D^{-1} P_L 
+ \sum_k \left( v_k, \delta_\zeta v_k \right) .
\end{equation}
The second term of the r.h.s. is discussed 
in \cite{geometrical-aspect} in analogy of the Berry connection.
It is refered as the measure term in 
\cite{abelian-chiral-gauge-theory,non-abelian-gauge-anomaly}.
This term may be expressed with a current as follows:
\begin{equation}
\sum_k \left( v_k, \delta_\zeta v_k \right) 
= -i a^4 \sum_x \zeta^a_\mu(x) j^a_\mu(x) .
\end{equation}

\subsubsection{Gauge invariant choice of the measure term 
in abelian chiral gauge theories}

It has been shown by L\"uscher in \cite{topology-and-axial-anomaly,
abelian-chiral-gauge-theory} that for anomaly-free abelian chiral 
gauge theories, the functional measure for 
the Weyl fermion fields can be constructed so that the gauge
invariance of the effective action is maintained exactly on the lattice.
If we consider a gauge transformation of the effective action
in an abelian chiral gauge theory, we obtain
\begin{equation}
\label{eq:effective-action-gauge-variation}
\delta_\omega  \ln \det M_{kj} = 
i \sum_x \omega(x) 
\left\{ {\rm tr} T \gamma_5\left(1-aD\right)(x,x)
                         - a^4 \partial_\mu^\ast j_\mu(x)
                       \right\}.
\end{equation}
In this respect, the result obtained by L\"uscher 
\cite{topology-and-axial-anomaly}, which is crucial for the gauge 
invariance, is that the anomaly associated with the chiral gauge current 
can be expressed in the following form:
\begin{equation}
\label{eq:abelian-gauge-anomaly}
{\rm tr} \gamma_5  \left(1- a D\right)  (x,x) 
= 
a^4\left\{
\frac{1}{32\pi^2}
\epsilon_{\mu\nu\rho\sigma} 
F_{\mu\nu}(x) F_{\rho\sigma}(x+\hat\mu+\hat\nu)  
+\partial^\ast_\mu \bar k_\mu(x) \right\} ,
\end{equation}
where $\bar k_\mu(x)$ is a gauge-invariant local current.
This is the consequence of the index theorem on the lattice 
\cite{index-theorem-at-finite-lattice}. 
Then it has been shown that
the basis $\{ v_j(x)\}$ can be chosen so that 
the current associated with the measure term is 
gauge-invariant and local and it satisfies the anomalous conservation law,
\begin{equation}
\label{eq:anomalous-conservation}
\partial_\mu^\ast j_\mu(x) 
= \frac{1}{a^4} {\rm tr} \gamma_5  \left(1- a D\right)  (x,x) 
= \partial^\ast_\mu \bar k_\mu(x),
\end{equation}
when the anomaly cancellation condition $\sum_i e_i^3 = 0$ is
satisfied.
The explicit form of the ansatz for the measure term is given by
\begin{eqnarray}
\label{eq:ansatz-measure-term}
-i a^4 \sum_k \zeta_\mu(x) j_\mu(x)
&=& 
- i \int^1_0 dt {\rm Tr}
 \left\{ 
   \hat P_R \left[ \partial_t \hat P_R , \delta_\zeta \hat P_R  \right] 
 \right\} 
\nonumber\\
&& - \int^1_0 dt \sum_x
  \left\{ 
   \zeta_{\, \mu}(x) \bar k_\mu(x) + A_\mu(x) \delta_\zeta \bar k_\mu(x) 
  \right\} , \nonumber\\
\end{eqnarray}
where $U_\mu(x,t)=\exp(i t A_\mu(x))$.

\subsection{Domain-wall fermion for vector-like theories}
\label{sec:domain-wall-fermion}

Neuberger's lattice Dirac operator has a close relation to 
the domain-wall fermion 
\cite{truncated-overlap,vranas-pauli-villars,kikukawa-noguchi}.
In a simplified formulation, 
the domain-wall fermion is defined by the five-dimensional
Wilson fermion with the Dirichlet boundary condition in
the fifth dimension.
With the Dirichlet boundary condition in the fifth direction,
\begin{equation}
\label{eq:Dirichlet-bc-in-5dim}
\left.  \psi_R (x,t) \right\vert_{t=-T}  = 0, \quad
\left.  \psi_L (x,t) \right\vert_{t=T+a_5}  = 0, \qquad
(T=N a_5), 
\end{equation}
the action of the domain-wall fermion is defined by
\begin{equation}
S_{\rm DW} = a_5 \sum_{t=-T+a_5}^{T}a^4 \sum_x \bar \psi(x,t) 
\left( D_{\rm 5w}-\frac{m_0}{a} \right) \psi(x,t), 
\quad (0 < m_0 < 2),
\end{equation}
where the gauge field is asuumed to be four-dimensional
\begin{equation}
  U_\mu(x,t) = U_\mu(x), \quad U_5(x,t) = 1 ,
\end{equation}
and the five-dimensional Wilson-Dirac operator $D_{\rm 5w}$ is 
defined as
\begin{eqnarray}
D_{\rm 5w}&=& 
\sum_{\mu=1}^4 
\left\{ 
\gamma_\mu \frac{1}{2}\left(\nabla_\mu - \nabla_\mu^\dagger \right) 
+\frac{a}{2} \nabla_\mu \nabla_\mu^\dagger\right\}
+ \gamma_5 \frac{1}{2}\left(\nabla_5 - \nabla_5^\dagger \right) 
+\frac{a_5}{2} \nabla_5 \nabla_5^\dagger . \nonumber\\
\end{eqnarray}

Due to its structure of the chiral hopping and the boundary condition 
in the fifth dimension, a single light Dirac fermion can emerge in the 
spectrum. This light fermion can be probed suitably
by the field variables at the boundary of the fifth dimension,
which are referred as $q(x)$ and $\bar q(x)$ by Furman and Shamir.
\begin{equation}
q(x)= \psi_L(x,-T+a_5) + \psi_R(x,T), \qquad
\bar q(x)= \bar \psi_L(x,-T+1) + \bar \psi_R(x,T) .
\end{equation}
In fact, the propagator of the light fermion field
can be expressed in terms 
of the effective Dirac operator 
\cite{truncated-overlap,kikukawa-noguchi}:
\begin{equation}
\label{eq:light-fermion-propagator}
\left\langle q(x) \bar q(y) \right\rangle 
= \frac{1}{a^4} 
\left( \frac{1}{a} { D_{\rm eff}^{(T)}}^{-1}-\delta(x,y) \right) ,
\end{equation}
where 
\begin{equation}
\label{eq:effective-Dirac-operator-finiteN}
D_{\rm eff}^{(T)} = \frac{1}{2a}\left(1 + \gamma_5 \tanh 
T \widetilde H \right) , \qquad (T=N a_5).
\end{equation}
$\widetilde H$ is defined through the transfer matrix of the 
five-dimensional Wilson fermion
\begin{equation}
\label{eq:transfer-matrix}
e^{- a_5 \tilde H} 
= \left(
\begin{array}{cc} \frac{1}{B} & - \frac{1}{B} C \\
                 -C^\dagger \frac{1}{B} 
& B + C^\dagger \frac{1}{B} C 
\end{array} \right),
\end{equation}
where
\footnote{
In this expression,
the positivity of $B$
is required for the transfer matrix to be defined consistently.
It is assured when $ 0 < \frac{a_5}{a} m_0  < 1$ . 
}
\begin{eqnarray}
\label{eq:operator-C}
  C &=& a_5 \, \sigma_\mu \, 
\frac{1}{2}\left(\nabla_\mu+\nabla_\mu^\ast\right) , \\
\label{eq:operator-B}
 B &=& 1 + a_5 \left( 
-\frac{a}{2} \nabla_\mu\nabla_\mu^\ast - \frac{m_0}{a} 
\right) .
\end{eqnarray}
The limit $T \rightarrow \infty$ is defined well
as long as $\widetilde H^2 > 0 $.
The effective Dirac operator 
Eq.~(\ref{eq:effective-Dirac-operator-finiteN}) then
reduces to Neuberger's lattice Dirac operator using $\widetilde H$,
\begin{equation}
\label{eq:effective-Dirac-operator}
D_{\rm eff} 
= 
\frac{1}{2a}
\left(1 + \gamma_5 \frac{\widetilde H}{\sqrt{\widetilde H^2}}\right),
\end{equation}
and turns out to satisfy the Ginsparg-Wilson relation.

It is useful to note that the effective Dirac operator admits 
the following representation \cite{kikukawa-noguchi,locality-in-dwf}:
\begin{eqnarray}
\label{eq:effective-Dirac-operator-Intro}
a D_{\rm eff}^{(T)} 
&=&  1- 
P_R \left\{ a_5 \left( \overline{D}_{\rm 5w}- \frac{m_0}{a}\right) 
    \right\}^{-1}_{T,T} P_L 
\nonumber\\
&& \quad 
- P_L \left\{ a_5 \left(\overline{D}_{\rm 5w}- \frac{m_0}{a}\right)
    \right\}^{-1}_{-T+a_5,-T+a_5} P_R 
\nonumber\\
&& \quad 
- P_R \left\{ a_5 \left(\overline{D}_{\rm 5w}- \frac{m_0}{a}\right)
      \right\}^{-1}_{T,-T+a_5} P_R
\nonumber\\
&& \quad 
- P_L \left\{ a_5 \left(\overline{D}_{\rm 5w}- \frac{m_0}{a}\right)
      \right\}^{-1}_{-T+a_5,T} P_L ,
\nonumber\\
\end{eqnarray}
where $\overline{D}_{\rm 5w}$ is  the five-dimensional 
Wilson-Dirac operator {\em with the anti-periodic boundary condition}
in the fifth-dimension. Its inverse may be expressed as
\begin{equation}
\left\{a_5  \left(\overline{D}_{\rm 5w}- \frac{m_0}{a}\right)
\right\}^{-1}_{st}  
= \frac{1}{2N} \sum_p 
\frac{e^{i p (s-t)}}
{i \gamma_5 \sin p a_5 + 1-\cos p a_5
        + a_5 \left(D_{\rm w}- \frac{m_0}{a}\right) }.
\end{equation}
The summation is taken over the discrete 
momenta ${p=\frac{\pi}{N a_5}(k-\frac{1}{2})}$ $(k=1,2,\cdots,2N)$ and
in the limit $N\rightarrow\infty$ it reduces to the continuous integral.

From this representation, it is rather clear that the effective 
Dirac operator can be defined consistently if the five-dimensional 
Wilson-Dirac operator with the anti-periodic boundary condition is 
not singular and invertible for all $T$.
In this respect, we should note that 
the lower bound on the square of the five-dimensional 
Wilson-Dirac operator is related closely 
to that on the square of the four-dimensional 
Wilson-Dirac operator \cite{locality-of-overlap-D,bound-neuberger},
because the gauge field is four-dimensional.
In fact, the same lower bound can be set for the class 
of gauge fields with small lattice field strength which satisfies
the bound Eq.~(\ref{eq:smooth-gauge-field}).
\begin{equation}
\left\| a^2 \left(\overline{D}_{\rm 5w}- \frac{m_0}{a}\right)^\dagger 
            \left(\overline{D}_{\rm 5w}- \frac{m_0}{a}\right)\right\|
\ge \,  \left\{ (1- 30 \epsilon)^{\frac{1}{2}}-|1-m_0| \right\}^2 .
\end{equation}

\section{A lattice implementation of the $\eta$-invariant}
\label{sec:implementation}
\reseteqnum

\subsection{The $\eta$-invariant, the five-dimensional massless fermion
on the lattice and the chiral domain-wall fermion}

In the continuum theory, the $\eta$-invariant can be defined 
through the complex phase of the determinant of a five-dimensional 
massless Dirac fermion in the Pauli-Villars regularization 
\cite{eta-invariant,anomaly-in-odd-dimensions}.  
In order to implement the $\eta$-invariant on the lattice, we may 
consider a five-dimensional massless Dirac fermion which is formulated 
on the lattice using the five-dimensional overlap Dirac operator 
\cite{overlap-in-odd-dimensions}. 
As shown in \cite{overlap-in-odd-dimensions}, in the overlap formalism 
\cite{overlap}, a five-dimensional massless Dirac fermion can be described
gauge invariantly by the five-dimensional overlap Dirac operator:
\begin{equation}
\label{eq:five-dimensional-massless-Dirac-fermion}
  S_{\rm 5dim}= a_5 a^4 \sum_{x,t} \bar \psi(x,t) 
\left( 1+ X_5 \frac{1}{\sqrt{X_5^\dagger X_5}}\right) 
\psi(x,t) ,
\end{equation}
where $X_5$ denotes the five-dimensional Wilson-Dirac operator 
$D_{\rm 5w}$ with a negative mass:
\begin{equation}
  X_5 = D_{\rm 5w} -\frac{m_0}{a} \qquad (0 < m_0 < 2).
\end{equation}
Then we can consider a lattice implementation of the $\eta$-invariant 
using the complex phase of the determinant
of the five-dimensional massless Dirac fermion which couples 
to a certain interpolating five-dimensional lattice gauge field:
\begin{equation}
\label{eq:lattice-implementation-eta-overlap}
\frac{\pi}{2} \, \overline{\eta} \, \, \simeq \, \, 
{\rm Im}  
\ln {\det}
\left( 1+ X_5 \frac{1}{\sqrt{X_5^\dagger X_5}}\right) .
\end{equation}
Since it holds that \cite{overlap-in-odd-dimensions}
\footnote{
If we use an abbreviation as
$  V=X_5 \frac{1}{\sqrt{X_5^\dagger X_5}} $, we have $V^\dagger V=1$.
Then, the complex phase of the partition function of the 
five-dimensional massless Dirac fermion can be evaluated as
\begin{equation}
 {\rm Im} \ln {\det}_{(T)}\left(1+V\right)
=\frac{1}{2i}\ln {\det}_{(T)} \frac{\left(1+V\right)}
                                   {\left(1+V^\dagger\right)} 
=\frac{1}{2i}\ln {\det}_{(T)} \, \frac{1}{V^\dagger}  
=\frac{1}{2}{\rm Im} \ln {\det}_{(T)} \, V .
\end{equation}
Since $\sqrt{X_5^\dagger X_5}$ is hermitian, it does not
contribute to the complex phase of the partition function and
can be neglected. 
Thus we obtain Eq.~(\ref{eq:5dim-overlap-to-domain-wall-fermion}).
}
\begin{equation}
\label{eq:5dim-overlap-to-domain-wall-fermion}
{\rm Im} \, \ln \, \det 
\left( 1+ X_5 \frac{1}{\sqrt{X_5^\dagger X_5}}\right) 
= \frac{1}{2} {\rm Im} \, \ln \, \det X_5 ,
\end{equation}
the above implementation can also be written as 
\begin{equation}
\label{eq:lattice-implementation-eta-domain-wall}
\pi \, \overline{\eta} \simeq
{\rm Im} \, \ln \, {\det}
\left( D_{\rm 5w} -\frac{m_0}{a} \right). 
\end{equation}

These consideration suggests that the $\eta$-invariant can be 
implemented on the lattice through the complex phase of 
the determinant of the simplified domain-wall fermion 
which couples to the interpolating five-dimensional gauge field.
Namely, we first impose the Dirichlet boundary condition 
in the fifth dimension and then let the extent of the fifth dimension
go to infinity.
\begin{equation}
\label{eq:lattice-implementation-eta-domain-wall-limit}
\pi \, \overline{\eta} \simeq
\lim_{T \rightarrow \infty} 
{\rm Im} \, \ln \, {\det}
\left( D_{\rm 5w(T)} -\frac{m_0}{a} \right). 
\end{equation}
$D_{\rm 5w(T)}$ stands for the Wilson-Dirac operator which is
subject to the Dirichlet boundary condition at $t=-T+a_5$ and
$t=T$.\footnote{
With this boundary condition, the difference operator in the fifth 
direction 
\begin{equation}
\partial_5 = \frac{1}{a_5}
\left(\delta_{x+\hat 5,y}-\delta_{x,y}\right) 
\end{equation}
may be expressed by $2N \times 2N$ matrix ($T=N a_5$) as follows:
\begin{equation}
\partial_{5(T)} = \frac{1}{a_5}\left(\begin{array}{cccccc}  
-1 & 1 & 0 & 0 & 0 & 0  \\
 0 & -1& 1 & 0 & 0 & 0  \\
 0 & 0 & -1 & 1 & 0 & 0  \\
 0 & 0 & 0 & -1 & 1 & 0  \\
 0 & 0 & 0 & 0  & -1 & 1  \\
 0 & 0 & 0 & 0 & 0 & -1   \end{array} \right)  \quad (N=3) .
\end{equation}
The subscript $(T)$ denotes the fact that the difference operator is 
implemented by a finite matrix, taking account of the Dirichlet 
boundary condition in the fifth direction. }
Because of the interpolation in the fifth direction, 
the chiral modes at the boundary $t=-T+a_5$ and $t=T$
couple to different four-dimensional gauge fields. 
This difference is expected to reprodue
the difference of the complex phase of the effective action for
the chiral Ginsparg-Wilson fermion, 
since the chiral modes at the boundaries are described 
by the effective Dirac operator satisfying the Ginsparg-Wilson relation.
Our proposal then can be regarded as a lattice realization 
of the argument given by Kaplan and Shmaltz in 
\cite{domain-wall-ferimon-and-eta-invariant}, 
using the simplified domain-wall fermion of Shamir 
\cite{boundary-fermion}.

\subsection{Smooth interpolation of four-dimensional gauge fields}

In order to realize the smooth interpolation of the four-dimensional 
lattice gauge fields, we need to choose carefully the interpolating
five-dimensional gauge field.
For this purpose, we first consider the five-dimensional lattice theory 
Eq.~(\ref{eq:lattice-implementation-eta-domain-wall-limit})
and then take the continuum limit in the fifth 
dimension, $a_5 \rightarrow 0$.

To prepare the interpolating gauge fields on the five-dimensional lattice, 
let us consider two four-dimensional gauge fields.
\begin{equation}
U_\mu(x)= e^{i A_\mu(x)}, \qquad 
U^0_\mu(x)= e^{i A^0_\mu(x)}.
\end{equation}
We assume that both gauge fields are smooth enough to 
satisfies the bound Eq.~(\ref{eq:smooth-gauge-field})
to make Neuberger's Dirac operator defined well and 
local \cite{locality-of-overlap-D}. 
We also assume that both gauge fields belong to the same 
topological sector in which the topological charge defined 
through Neuberger's Dirac operator 
\begin{equation}
Q= - a {\rm Tr} \gamma_5 D 
 = - \frac{1}{2} {\rm Tr} \frac{H}{\sqrt{ H^2}} 
\end{equation}
are equal.

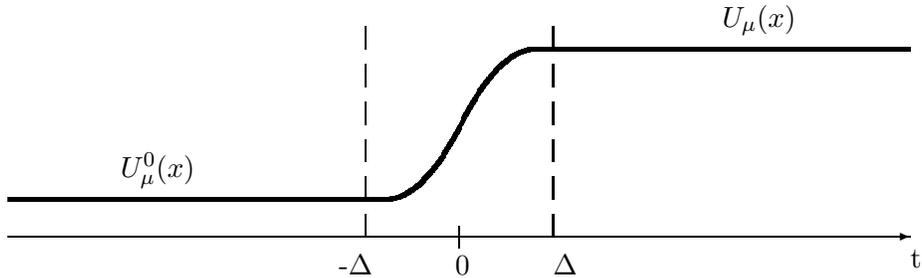
\begin{figure}[htbp]
  \begin{center}
{\unitlength 1mm
\begin{picture}(160,40)
\put(0,0){\vector(1,0){120}}
\put(60,-1.5){\line(0,1){3}}
\put(59.5,-5){0}
\put(120,-4){t}

{\linethickness{0.5mm}
\put(0,5){\line(1,0){50}}


\put(70,25){\line(1,0){50}}
\qbezier(50,5)(55,5)(60,15)
\qbezier(60,15)(65,25)(70,25)

}

\multiput(47.5,0)(0,5){6}{\line(0,1){3}}
\put(44,-5){-$\Delta$}
\multiput(72.5,0)(0,5){6}{\line(0,1){3}}
\put(72.5,-5){$\Delta$}

\put(15,8){$U^0_\mu(x)$}

\put(95,28){$U_\mu(x)$}

\end{picture}
}
    \caption{Interpolating five-dimensional gauge field}
    \label{fig:continuous-interpolation}
  \end{center}
\end{figure}

Then we consider a five-dimensional gauge field 
interpolating two four-dimensional gauge fields,
$U^0_\mu(x)$ and  $U_\mu(x)$, along the fifth coordinate $t$
which is for the first time regarded as a continuous coordinate.
\begin{equation}
  U_\mu(x,t)  \qquad t \in R.
\end{equation}
We assume that it is possible to choose such an interpolation 
without breaking the bound Eq.~(\ref{eq:smooth-gauge-field}) and 
changing the topological property of the gauge fields:
\begin{equation}
\label{eq:smooth-gauge-field-all-t}
\left\Arrowvert  1-U_{\mu\nu}(x,t) \right\Arrowvert 
< \epsilon, 
\qquad 
\epsilon < \frac{1}{30}\left\{1-|1-m_0|^2\right\}, \quad t \in R.
\end{equation}
We also assume that the interpolating region has a finite interval, 
$ t \in [-\Delta, \Delta]$:
When $t <  -\Delta$, it coincides with the 
four-dimensional gauge field $U^0_\mu(x)$,
\begin{equation}
U_\mu(x,t) \ \larrow_{ t < -\Delta} \  U^0_\mu(x).
\end{equation}
When $ t > \Delta$, it coincides with the other four-dimensional
gauge field $U_\mu(x)$:
\begin{equation}
U_\mu(x,t) \ \larrow_{ t > +\Delta} \  U_\mu(x).
\end{equation}
This defines one parameter family of the interpolating 
five-dimensional gauge fields. 
See Figure~\ref{fig:continuous-interpolation}.

\begin{figure}[htbp]
  \begin{center}
{\unitlength 1mm
\begin{picture}(160,40)
\put(20,0){\line(1,0){85}}
\put(60,-1.5){\line(0,1){3}}
\put(59.5,-5){0}

\multiput(20,0)(5,0){18}{\circle*{1}}

\put(20,0){\line(0,1){30}}
\put(10,-5){$-T+a_5$}
\put(105,0){\line(0,1){30}}
\put(105,-5){$T(=Na_5)$}

{\linethickness{0.5mm}

\put(20,5){\line(1,0){30}}

\put(70,25){\line(1,0){35}}
\qbezier(50,5)(55,5)(60,15)
\qbezier(60,15)(65,25)(70,25)

}
\multiput(0,5)(5,0){4}{\line(1,0){3}}
\multiput(105,25)(5,0){4}{\line(1,0){3}}

\multiput(47.5,0)(0,5){6}{\line(0,1){3}}
\put(44,-5){-$\Delta$}
\multiput(72.5,0)(0,5){6}{\line(0,1){3}}
\put(72.5,-5){$\Delta$}

\put(22,8){$U^0_\mu(x)$}
\put(90,28){$U_\mu(x;u)$}

\end{picture}
}
    \caption{Interpolating five-dimensional gauge field on the lattice}
    \label{fig:lattice-interpolation}
  \end{center}
\end{figure}
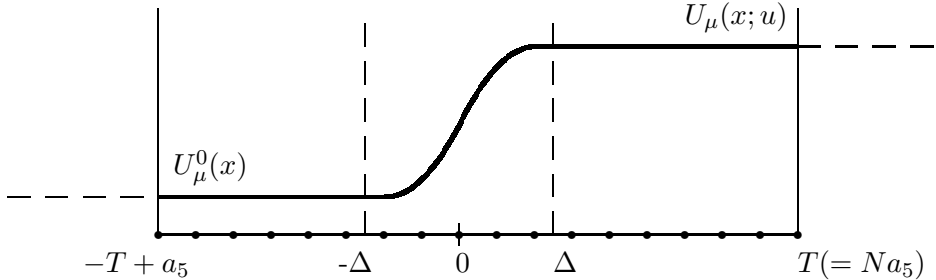

We then map the continuum interpolations to a discrete fifth dimensional
lattice space so that 
$\Delta <  T $. (Figure~\ref{fig:lattice-interpolation})
\begin{equation}
  U_\mu(x,t)  \qquad t=n_5 a_5, \quad n_5 \in Z.
\end{equation}
This interpolating five-dimensional lattice gauge field is to couple 
to the domain-wall fermion of 
Eq.~(\ref{eq:lattice-implementation-eta-domain-wall-limit}).
In order to recover the smooth interpolation of the two 
four-dimensional lattice gauge fields, we need 
to take both the infinite extent limit $T \rightarrow \infty$ 
and the continuum limit $a_5 \rightarrow 0$ of the fifth
dimension, keeping  $ \Delta \ll T $.

\subsection{Inverse five-dimensional Wilson-Dirac operators}

For technical reasons, we also require that 
the five-dimensional Wilson-Dirac operator
which couples to the five-dimensional interpolating lattice gauge 
fields does not have zero mode and is invertible. 
For this, we asuume the following bound on the $5-\mu$ 
plaquette,\footnote{
Y.K. is grateful to D.B.~Kaplan for discussions and suggestions
on this point.}
\begin{equation}
\label{eq:smooth-gauge-field-5-dim}
\left\Arrowvert  1-U_{5\mu}(x,t) \right\Arrowvert 
< \left(\frac{a_5}{a}\right) \epsilon_5, 
\qquad 
30 \epsilon + 20 \epsilon_5 < \left\{1-|1-m_0|^2\right\}.
\end{equation}
Note that since we can estimate the size of the 
$5-\mu$ plaquette as 
\begin{equation}
\| 1-U_{5\mu}(x,t) \| 
\simeq \frac{a_5}{\Delta} \| U^0_\mu(x) - U^1_\mu(x) \|,
\end{equation}
and
\begin{equation}
\epsilon_5 \simeq \frac{a}{\Delta}   \| U^0_\mu(x) - U^1_\mu(x) \|,
\end{equation}
this bound holds true as long as we choose $\Delta /a$ large enough.
\footnote{The bound on the five-dimensional plaquette variables
can be regarded as a sufficient condition for the existence 
of chiral fermions on the boundary walls.
In this respect, we note that 
in the waveguide approach of domain wall fermion 
\cite{waveguide-kaplan,waveguide-jansen-etal},
$\Delta /a$ is set to unity and the bound on the 
five-dimensional plaquette variables is not satisfied in general.
}
Then the five-dimensional Wilson-Dirac operator is bounded from
below by a positive constant,
\begin{equation}
\left\| a^2 \left(D_{\rm 5w}- \frac{m_0}{a}\right)^\dagger 
            \left(D_{\rm 5w}- \frac{m_0}{a}\right)\right\|
\ge \,  
\left\{ (1- 30 \epsilon - 20 \epsilon_5)^{\frac{1}{2}}-|1-m_0| \right\}^2 .
\end{equation}

Given the potitive lower and upper bounds for the five-dimensional
Wilson-Dirac operator, 
\begin{equation}
\tilde \alpha \le
\left\| a^2 \left(D_{\rm 5w}- \frac{m_0}{a}\right)^\dagger 
            \left(D_{\rm 5w}- \frac{m_0}{a}\right)\right\|
\le \tilde \beta,  
\end{equation}
it follows that 
the inverse five-dimensional Wilson-Dirac operator decays
exponentially at large distance in the fifth 
dimension \cite{locality-in-dwf}:
\begin{equation}
\label{eq:exponetial-bound-on-5dim-D}
\left\| \left\{a^2 D_{\rm 5w}^\dagger  
D_{\rm 5w}\right\}^{-1}(x,s;y,t) 
\right\| \le  \, 
C \, 
\exp \left\{ - \frac{\tilde \theta}{2}d_5(x,s;y,t)\right\} ,
\end{equation}
where
\begin{equation}
C=\frac{4 t}{\tilde \beta-\tilde \alpha} 
\left( 
\frac{1}{1-t} \frac{d_5(x,s;y,t)}{2} + \frac{t}{(1-t)^2} 
\right),
\end{equation}
\begin{equation}
t= e^{-\tilde \theta}, \qquad \cosh 
\tilde \theta = \frac{\tilde \beta + \tilde \alpha}
              {\tilde \beta - \tilde \alpha} ,
\end{equation}
and $d_5(x,s;y,t)=|x-y|/a +|s-t|/a_5$.

The similar property holds true for the five-dimensional Wilson-Dirac
operator which is subject to the Dirichlet boundary condition 
in the fifth direction. In order to see this, 
we note the following identity which holds for $s,t \in [-T+a_5,T]$:
\begin{eqnarray}
\label{eq:5d-Dirac-operator-inverse-diff}
&&
 \frac{1}{ D_{5w(T)} -\frac{m_0}{a} }\,(s,x;t,y)
-\frac{1}{ D_{5w} -\frac{m_0}{a} }\,(s,x;t,y)  \nonumber\\
&& \qquad\qquad \qquad\quad 
=\frac{1}{ D_{5w(T)} -\frac{m_0}{a} } \, 
V_{(-T+a_5;T)}  \,
 \frac{1}{ D_{5w} -\frac{m_0}{a} } \, (s,x;t,y) , \nonumber\\
\end{eqnarray}
where 
\begin{eqnarray}
\label{eq:surface-interaction}
V_{(-T+a_5;T)} &=&   
\frac{1}{a_5}\left\{ - P_L \delta_{s,-T}\delta_{t,-T+a_5} 
- P_R \delta_{s,-T+a_5}\delta_{_t,-T} 
\right. 
\nonumber\\
&& \qquad\qquad
\left.
- P_L \delta_{s,T}\delta_{t,T+a_5} 
- P_R \delta_{s,T+a_5}\delta_{t,T} \right\}.
\end{eqnarray}
In this identity, the Dirichlet boundary condition 
at $t=-T+a_5$ and $t=T$ is implemented in the infinite extent of the 
fifth dimension by adding the surface interaction 
term \cite{symanzik}.
The derivation of this identity is given in 
appendix~\ref{app:5d-Dirac-operator-inverse-diff}.
By setting $t=T$, we obtain
\begin{eqnarray}
\label{eq:5d-Dirac-operator-inverse-diff-2}
&&
 \frac{1}{ D_{5w(T)} -\frac{m_0}{a} }\,(s,T)
\left(1+P_L  \,
 \frac{1}{ D_{5w} -\frac{m_0}{a} } \, (T+a_5,T) \right) \nonumber\\
&& 
=
\frac{1}{ D_{5w} -\frac{m_0}{a} }\,(s,T)  \nonumber\\
&&
\quad -\frac{1}{ D_{5w(T)} -\frac{m_0}{a} }(s,-T+a_5) \, 
P_R \,
 \frac{1}{ D_{5w} -\frac{m_0}{a} } \, (-T,T) . \nonumber\\
\end{eqnarray}
We may assume that the correlator
\begin{equation}
\left( D_{5w} -\frac{m_0}{a} \right)^{-1}(T+a_5,T)  
\end{equation}
has a finite limit when $T \rightarrow \infty$. 
Then 
we can infer that the inverse of $D_{5w(T)} -\frac{m_0}{a}$ should 
decay exponentially (up to power corrections in $T$) at large distance
with the same exponent as Eq.~(\ref{eq:exponetial-bound-on-5dim-D})
for $D_{5w} -\frac{m_0}{a}$. Otherwise, it would contradict with
Eq.~(\ref{eq:5d-Dirac-operator-inverse-diff-2}).
Therefore we obtain 
the following bound in the limit $T \rightarrow \infty$ and
$|s-T| \rightarrow \infty$,
\begin{eqnarray}
\label{eq:exponetial-bound-on-5dim-D-Dirichlet}
&&
\left \|  \frac{1}{ D_{5w(T)} -\frac{m_0}{a} }\,(x,s;y,T) \right\|
\, \le \, 
C^\prime \, (T/a_5)^m
\exp \left\{ - \frac{\tilde \theta}{2a_5} |s-T|\right\} ,
\nonumber\\
&& \qquad \qquad \qquad \qquad \qquad 
   \qquad \qquad \qquad \qquad 
\left( |s-T| \rightarrow \infty \right),
\end{eqnarray}
with a positive constant $C^\prime$ and a positive integer $m$.

\subsection{Definition of the $\eta$-invariant on the lattice}

In summary, we consider the following lattice implementation of the
$\eta$-invariant.
\begin{equation}
\label{eq:lattice-implementation-eta-domain-wall-limit-limit}
\pi \, \overline{\eta} \equiv
\lim_{a_5 \rightarrow 0} 
\lim_{T \rightarrow \infty} 
{\rm Im} \, \ln \, {\det}
\left( D_{\rm 5w(T)} -\frac{m_0}{a} \right). 
\end{equation}
It utilizes the complex phase of 
the determinant of the simplified domain-wall fermion 
which couples to the interpolating five-dimensional gauge field,
\begin{equation}
S_{\rm DW} = a_5 \sum_{t=-T+a_5}^{T}a^4 \sum_x \bar \psi(x,t) 
\left( D_{\rm 5w}-\frac{m_0}{a} \right) \psi(x,t), 
\qquad (0 < m_0 < 2),
\end{equation}
with the Dirichlet boundary condition in the fifth direction,
\begin{equation}
  \psi_R (x,t) \left\vert_{t=-T} \right. = 0, \qquad
  \psi_L (x,t) \left\vert_{t=T+ a_5} \right. = 0.
\end{equation}
$D_{\rm 5w(T)}$ stands for the five-dimensional Wilson-Dirac operator 
$D_{\rm 5w}$ which is subject to the Dirichlet boundary condition.
The five-dimemsional lattice gauge field ``interpolates''
two four-dimensional lattice gauge field with same topological charge
in a finite region $t \in [-\Delta,\Delta]$.
It satisfies the bounds on the plaquette variables as 
\begin{eqnarray}
&& \left\Arrowvert  1-U_{\mu\nu}(x,t) \right\Arrowvert 
< \epsilon, 
\\
&&
\left\Arrowvert  1-U_{5\mu}(x,t) \right\Arrowvert 
< \left(\frac{a_5}{a}\right) \epsilon_5, 
\quad 
30 \epsilon + 20 \epsilon_5 < \left\{1-|1-m_0|^2\right\}.
\end{eqnarray}
and has a smooth continuum limit in $a_5 \rightarrow 0$.

\section{The variation of $ \overline{\eta} $ with respect to gauge field}
\label{sec:variation}
\reseteqnum

Following the analysis in the continuum theory\cite{eta-invariant}, 
we next examine the variation of $\overline{\eta}$ with respect to 
the gauge field. 
For this purpose, we introduce another parameter 
$u \in [0,1]$ so that it parametrizes the gauge field configuration
at $t \ge +\Delta$ from $U^0_\mu(x)$ to $U_\mu(x)$.
(Figure~\ref{fig:parameter-u})
\begin{equation}
\label{eq:one-family-4dim-gauge-fields}
U_\mu(x;u) , \qquad u \in [0,1] .  
\end{equation}
\begin{equation}
U_\mu(x;u=0) = U^0_\mu(x), \quad
U_\mu(x;u=1) = U_\mu(x) .
\end{equation}
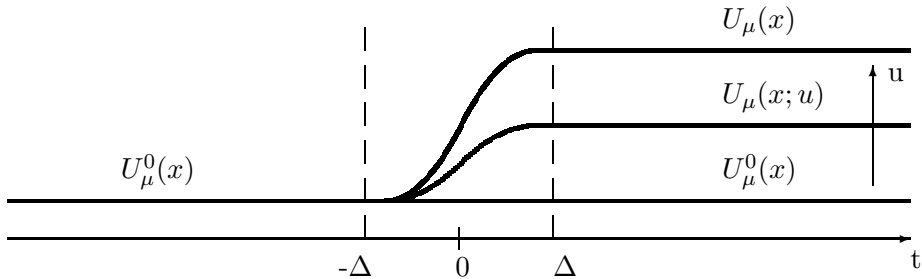
\begin{figure}[htbp]
  \begin{center}
    
{\unitlength 1mm
\begin{picture}(160,40)
\put(0,0){\vector(1,0){120}}
\put(60,-1.5){\line(0,1){3}}
\put(59.5,-5){0}
\put(120,-4){t}

{\linethickness{0.5mm}
\put(0,5){\line(1,0){50}}

\put(50,5){\line(1,0){70}}

\put(70,25){\line(1,0){50}}
\qbezier(50,5)(55,5)(60,15)
\qbezier(60,15)(65,25)(70,25)

\put(70,15){\line(1,0){50}}
\qbezier(50,5)(55,5)(60,10)
\qbezier(60,10)(65,15)(70,15)
}

\multiput(47.5,0)(0,5){6}{\line(0,1){3}}
\put(44,-5){-$\Delta$}
\multiput(72.5,0)(0,5){6}{\line(0,1){3}}
\put(72.5,-5){$\Delta$}

\put(15,8){$U^0_\mu(x)$}

\put(95,28){$U_\mu(x)$}
\put(95,18){$U_\mu(x;u)$}
\put(95,8){$U^0_\mu(x)$}

\put(115,7){\vector(0,1){16}}
\put(117,21){u}

\end{picture}
}
    \caption{Parameter $u$}
    \label{fig:parameter-u}
  \end{center}
\end{figure}

\subsection{Summary of result}
Before going into technical details, we first 
summarize our result.
The variation of $\overline{\eta}$ with respect to $u$ can be
written as the sum of two contributions as follows:
\begin{eqnarray}
\label{eq:variation-eta-result}
\frac{d}{du}\, \overline{\eta}\left[U_\mu(x,t;u)\right]
&=&
\lim_{a_5 \rightarrow 0} \lim_{T \rightarrow \infty} 
\frac{1}{\pi} {\rm Im} {\rm Tr}_{(T)} \, 
\, \frac{d}{du} D_{\rm 5w(\infty)} 
\, \frac{1}{ D_{\rm 5w(\infty)} -\frac{m_0}{a}}
  \nonumber\\
&& 
+ \frac{1}{\pi} {\rm Im} \,
{\rm Tr}_x \, P_L\, \frac{d}{du} D \, \frac{1}{D} .
\end{eqnarray}
The first one is the bulk five-dimensional contribution, 
which depends on the whole interpolating five-dimensional gauge fields.
The second one is the contribution from the boundaries at $t=-T+a_5$ 
and $t=T$, which depends only on the boundary values of the interpolating 
five-dimensional gauge fields.

Remarkably, the surface contribution is expressed by the covariant 
chiral gauge current associated with
Neuberger's Dirac operator. 
This surface term can be related to the imaginary part 
of the effective action for the chiral Ginsparg-Wilson fermion
which is defined with Neuberger's Dirac operator \cite{overlap,
abelian-chiral-gauge-theory,suzuki}, as we will see later.

The bulk contribution
reproduces the Chern-Simons term in the classical continuum limit. 
First of all, this term is 
a local functional of $U_\mu(x,t)$. This follows from 
the property of the inverse five-dimensional Wilson-Dirac operator given
by Eq.~(\ref{eq:exponetial-bound-on-5dim-D}), which 
becomes small exponentially at large distance.
Secondly, using a plain wave basis on the lattice, 
as in the continuum analysis, it can be evaluated as
\begin{eqnarray}
&&  \lim_{a_5 \rightarrow 0} \lim_{T \rightarrow \infty} 
{\rm Im} {\rm Tr}_{(T)} \, 
\, \frac{d}{du} D_{\rm 5w(\infty)} 
\, \frac{1}{ D_{\rm 5w(\infty)} -\frac{m_0}{a}}
\nonumber\\
&& \ce_{ a,a_5\rightarrow 0}
-\lim_{T' \rightarrow \infty} 
\int d^4 x \int^{T'}_{-T'} dt \, 
\frac{1}{32\pi^2} \, \epsilon_{\mu MNKL} 
{\rm tr}\left\{ \frac{d}{du} A_\mu \, F_{MN} F_{KL}\right\}(x,t;u) .
\nonumber\\
\end{eqnarray}
This quantity can be written as the variation of the Chern-Simons 
term, up to the local current of Bardeen and Zumino
\cite{bardeen-zumino} which plays the role to 
translate the covariant gauge current from the surface 
contribution to the consistent one \cite{eta-invariant}.
Thus the result in the continuum theory \cite{eta-invariant}
is completely reproduced by our lattice implementation 
of the $\eta$-invariant in the classical continuum 
limit.\footnote{We are considering the 
effective action for the right-handed Weyl fermions.}

The coefficient of the Chern-Simons term in this calculation 
is given by the topological number associated with the free
five-dimensional Wilson-Dirac operator,
\begin{eqnarray}
c&=& \frac{1}{5!} \int^\pi_{-\pi} \frac{d^5 k}{(2\pi)^5}  \,
\epsilon_{MNIJK} 
{\rm Tr} \left\{ 
\left(\partial_M S^{-1} S\right)
\left(\partial_N S^{-1} S\right)
\times  \right. \nonumber\\
&& \qquad\qquad\qquad\qquad \left. 
\left(\partial_I S^{-1} S\right) 
\left(\partial_J S^{-1} S\right)
\left(\partial_K S^{-1} S\right)
\right\} (k) \nonumber\\
&=& \frac{i}{8\pi^2}, 
\end{eqnarray}
where $S(k)$ is the free propagator of five-dimensional Wilson-Dirac 
fermion.
\begin{equation}
 S^{-1}(k)
=\sum_{M=1}^5
 \left( i \gamma_M \sin k_M + 2 \sin^2 \frac{k_M}{2} \right) -m_0 
\quad (0 < m_0 < 2) .
\end{equation}
This result is consistent with the previous calculation of 
the Chern-Simons current by Golterman, Jansen and 
Kaplan \cite{5dim-CS-in-domain-wall-fermion}.
The similar quantity in which the fifth momentum is continuous 
has appeared in the calculation of the axial anomaly 
\cite{kikukawa-yamada} of 
the Ginsparg-Wilson fermion defined with Neuberger's Dirac 
operator.\footnote{See 
\cite{3dim-CS} for the original calculations of the Chern-Simons term
induced from the Wilson-Dirac fermion in three dimensions.
For the detail analysis of the chiral Jaccobian of 
the Ginsparg-Wilson fermion, the authors refer the reader 
to \cite{lattice-chiral-jaccobian,chiu-anomaly}
} 

\subsection{Evaluation of $\frac{d}{du}\, \overline{\eta } $}

\subsubsection{Separation of bulk contribution}
Now we go into details how to evaluate the variation of $\overline{\eta}$.
From the continuum argument of \cite{eta-invariant},
we expect that the variation of $\overline{\eta}$ can be written as
the sum of two contributions. The first one is 
the bulk five-dimensional contribution, 
which should reproduce a part of the Chern-Simons term.
The second one is the contribution from the boundaries at $t=-T+a_5$ 
and $t=T$, which should be related to the effective action of the 
chiral fermion. In the context of the domain-wall fermion here, it
should be related to the chiral light modes at the boundaries.

By taking the variation of $\overline{\eta}$ with respect to $u$, 
we obtain
\begin{eqnarray}
\frac{d}{du}\, \overline{\eta }
&=& 
\label{eq:variation-eta}
\lim_{a_5 \rightarrow 0} \lim_{T \rightarrow \infty} 
\frac{1}{\pi} {\rm Im} {\rm Tr}_{(T)} \, \frac{d}{du} D_{\rm 5w(T)} \, 
\frac{1}{ D_{\rm 5w(T)} -\frac{m_0}{a}} .
\end{eqnarray}
Note that since $D_{5(T)}$ is defined 
in the finite interval of $[-T+a_5,T]$, taking account of 
the Dirichlet boundary condition, the cyclic property of the trace 
over the fifth dimension holds true here.

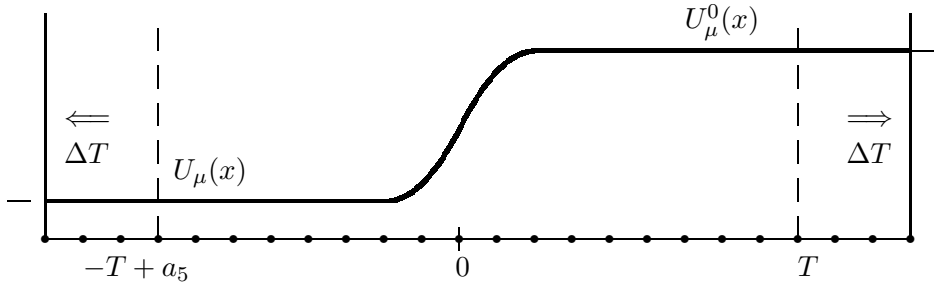
\begin{figure}[htbp]
  \begin{center}

{\unitlength 1mm
\begin{picture}(160,40)
\put(5,0){\line(1,0){115}}
\put(60,-1.5){\line(0,1){3}}
\put(59.5,-5){0}

\multiput(5,0)(5,0){24}{\circle*{1}}
\put(5,0){\line(0,1){30}}
\multiput(20,0)(0,5){6}{\line(0,1){3}}
\put(120,0){\line(0,1){30}}
\multiput(105,0)(0,5){6}{\line(0,1){3}}
\put(10,-5){$-T+a_5$}
\put(105,-5){$T$}
\put(7.5,15){$\Longleftarrow$}
\put(7.5,10){$\Delta T$}
\put(111.5,15){$\Longrightarrow$}
\put(111.5,10){$\Delta T$}

{\linethickness{0.5mm}
\put(5,5){\line(1,0){45}}
\put(70,25){\line(1,0){50}} 
\qbezier(50,5)(55,5)(60,15)
\qbezier(60,15)(65,25)(70,25)
}
\multiput(0,5)(5,0){3}{\line(1,0){3}}
\multiput(110,25)(5,0){3}{\line(1,0){3}}

\put(22,8){$U_\mu(x)$}
\put(90,28){$U^0_\mu(x)$}

\end{picture}
}
    \caption{Larger five-dimensional space}
    \label{fig:larger-5dim-space}
  \end{center}
\end{figure}

In order to separate the bulk five-dimensional contribution from
the boundary contribution, we note the following fact:
the bulk term comes from the interval $[-\Delta,\Delta]$ 
where the interpolating field is varing in $t$. 
Then it can also be evaluated from the five-dimensional Dirac 
fermion (the simplified domain-wall fermion) 
defined in a slightly larger five dimensional space
than $t \in [-T+a_5,T]$, say, 
$t \in [-T-\Delta T+a_5,T+\Delta T]$. 
(Figure~\ref{fig:larger-5dim-space})

\noindent
The inverse of the Dirac operator in this case
\begin{equation}
\label{eq:5d-Dirac-inverse-enlarged}
\frac{1}{D_{\rm 5w(T+\Delta T)} -\frac{m_0}{a}} \, (x,s;y,t)
\end{equation}
does not support the light chiral modes at the original boundaries 
$t=-T+a_5$ and $t=T$. If we would replace the inverse of the 
five-dimensional Dirac operator in Eq.~(\ref{eq:variation-eta}) by
Eq.~(\ref{eq:5d-Dirac-inverse-enlarged}),
then it could include only the bulk contribution. 
This consideration suggests the following separation:
\begin{eqnarray}
\label{eq:variation-eta-bulk-boundary}
&& \frac{1}{\pi} {\rm Im} 
{\rm Tr}_{(T)} \, 
\, \frac{d}{du} D_{\rm 5w(T)} \,
\frac{1}{ D_{\rm 5w(T)} -\frac{m_0}{a}} \nonumber\\
&=& 
\frac{1}{\pi} {\rm Im} 
{\rm Tr}_{(T)} \, 
\frac{d}{du} D_{\rm 5w(T)} \, 
\frac{1}{ D_{\rm 5w(T+\Delta T)} -\frac{m_0}{a} }  
\nonumber\\
&&
+\frac{1}{\pi} {\rm Im} 
{\rm Tr}_{(T)} \, 
\frac{d}{du} D_{\rm 5w(T)} \, \left(
 \frac{1}{D_{\rm 5w(T)} -\frac{m_0}{a}z }  
-\frac{1}{D_{\rm 5w(T+\Delta T)} -\frac{m_0}{a}} \right) . 
\nonumber\\
\end{eqnarray}
Note that we can let $\Delta T$ be infinity in this separation.

In order to see that the second term in the r.h.s. of 
Eq.~(\ref{eq:variation-eta-bulk-boundary}) is actually 
localized at the boundary, 
we note the following identity which holds for 
$s,t \in [-T+a_5,T]$:
\begin{eqnarray}
\label{eq:5d-Dirac-operator-inverse-diff-3}
&&
 \frac{1}{ D_{\rm 5w(T)} -\frac{m_0}{a} }\,(s,x;t,y)
-\frac{1}{ D_{\rm 5w (T+\Delta T)} -\frac{m_0}{a} }\,(s,x;t,y)  \nonumber\\
&& \qquad\qquad \qquad\quad 
=\frac{1}{ D_{\rm 5w(T)} -\frac{m_0}{a} } \, 
V_{(-T+a_5;T)}  \,
 \frac{1}{ D_{\rm 5w(T+\Delta T)} -\frac{m_0}{a} } \, (s,x;t,y) , \nonumber\\
\end{eqnarray}
where 
\begin{eqnarray}
\label{eq:surface-interaction-3}
V_{(-T+a_5;T)} &=& \frac{1}{a_5} \left\{  
- P_L \delta_{s,-T}\delta_{t,-T+a_5} 
- P_R \delta_{s,-T+a_5}\delta_{_t,-T} 
\right.
\nonumber\\
&& 
\left. \qquad \qquad
- P_L \delta_{s,T}\delta_{t,T+a_5} 
- P_R \delta_{s,T+a_5}\delta_{t,T} \right\}.
\end{eqnarray}
In this identity,\footnote{
In the limit $\Delta T \rightarrow \infty$, it reduces to 
Eq.~(\ref{eq:5d-Dirac-operator-inverse-diff}).}
the Dirichlet boundary condition 
at $t=-T+a_5$ and $t=T$, in the middle  of the enlarged extent 
of the fifth dimension $[-T-\Delta T+a_5,T+\Delta]$, is 
implemented by adding the surface interaction 
term \cite{symanzik}.
The derivation of this identity is given in 
appendix~\ref{app:5d-Dirac-operator-inverse-diff}.

Using this identity, the second term can be evaluated as follows:
\begin{eqnarray}
&& 
\frac{1}{\pi} {\rm Im} 
{\rm Tr}_{(T)} \, 
\frac{d}{du} D_{\rm 5w(T)} \left(
 \frac{1}{D_{\rm 5w(T)} -\frac{m_0}{a}z }  
-\frac{1}{D_{\rm 5w(T+\Delta T)} -\frac{m_0}{a}} \right)  \nonumber\\
&=&
\frac{1}{\pi} {\rm Im} 
{\rm Tr}_{(T)} \, 
\frac{d}{du} D_{\rm 5w(T)} \, 
\frac{1}{ D_{\rm 5w(T)} -\frac{m_0}{a} } V_{(-T+1,T)} 
\frac{1}{ D_{\rm 5w(T+\Delta T)} -\frac{m_0}{a} }  \nonumber\\
&=&
-\frac{1}{\pi} {\rm Im} 
{\rm Tr}_{(T)} \, 
\frac{d}{du} \left( \frac{1}{ D_{\rm 5w(T)} -\frac{m_0}{a} } \right)
V_{(-T+a_5,T)} 
\left( 1 - \frac{1}{ D_{\rm 5w(T+\Delta T)} -\frac{m_0}{a} }
V_{(-T+a_5,T)} \right) . \nonumber\\
\end{eqnarray}
Inserting the explicit expression of $V_{(-T+a_5;T)}$, we can 
evaluate it further to have
\begin{eqnarray}
\label{eq:boundary-contributions}
&=&\frac{1}{\pi} {\rm Im} {\rm Tr}_x \,
\frac{1}{a_5^2}
\left\{   
P_L \, \frac{d}{du} \left( \frac{1}{ D_{\rm 5w(T)} -\frac{m_0}{a} } \right)
{\scriptstyle (-T+a_5;-T+a_5)} \, 
P_R \frac{1}{ D_{\rm 5w(T+\Delta T)} -\frac{m_0}{a} } 
{\scriptstyle (-T;-T)} 
\right. \nonumber\\
&&
\qquad \qquad \qquad \left.
+
P_R \, \frac{d}{du} \left( \frac{1}{ D_{\rm 5w(T)} -\frac{m_0}{a} } \right)
{\scriptstyle (T;T)} \, 
P_L
\frac{1}{ D_{\rm 5w(T+\Delta T)} -\frac{m_0}{a} } 
{\scriptstyle (T+a_5;T+a_5)} 
\right. \nonumber\\
&&
\qquad \qquad \qquad \left.
+
P_L \, \frac{d}{du} \left( \frac{1}{ D_{\rm 5w(T)} -\frac{m_0}{a} } \right)
{\scriptstyle (-T+a_5;T)} \, 
P_R
\frac{1}{ D_{\rm 5w(T+\Delta T)} -\frac{m_0}{a} } 
{\scriptstyle (T;-T)} 
\right. \nonumber\\
&&
\qquad \qquad \qquad \left.
+
P_R \, \frac{d}{du} \left( \frac{1}{ D_{\rm 5w(T)} -\frac{m_0}{a} } \right)
{\scriptstyle (T;-T+a_5)} \, 
P_L
\frac{1}{ D_{\rm 5w(T+\Delta T)} -\frac{m_0}{a} } 
{\scriptstyle (-T;T+a_5)} 
\right\} . \nonumber\\
\end{eqnarray}
We can see that 
the first two terms are localized at the boundaries $t=-T+a_5$ and
$t=T$, respectively. The last two terms comes from the correlation 
between two boundaries. 

After letting $\Delta T$ go to infinity, 
we can see that the last two terms in Eq.~(\ref{eq:boundary-contributions})
vanish in the limit $T\rightarrow \infty$, because
the inverse of the five-dimensional Wilson-Dirac operator vanishes
exponentially for a large separation 
in the fifth dimension. 
See Eqs.~(\ref{eq:exponetial-bound-on-5dim-D}) and
(\ref{eq:exponetial-bound-on-5dim-D-Dirichlet}).

Therefore, we can write the variation of $\overline{\eta}$ as
\begin{equation}
  \frac{d}{du} \, \overline{\eta} = 
{\frac{d}{du} \,\overline{\eta}}_{\rm bulk}
+{\frac{d}{du} \, \overline{\eta}}_{\rm surf} ,
\end{equation}
\begin{eqnarray}
\label{eq:eta-variation-bulk}
{\frac{d}{du} \, \overline{\eta}}_{\rm bulk}
&=& \lim_{a_5 \rightarrow 0} \lim_{T \rightarrow \infty} 
\frac{1}{\pi} {\rm Im} 
{\rm Tr}_{(T)} \, 
\frac{d}{du} D_{\rm 5w} \frac{1}{ D_{\rm 5w(\infty)} -\frac{m_0}{a} }, \\
\label{eq:eta-variation-surf}
{\frac{d}{du}\, \overline{\eta}}_{\rm surf}  
&=&
\lim_{a_5 \rightarrow 0} \lim_{T \rightarrow \infty} 
\frac{1}{\pi} {\rm Im} \, {\rm Tr}_x \,
\frac{1}{a_5^2}
\left\{   \phantom{ \frac{d}{du}}
\right.   \nonumber \\
&& \left. 
P_L \frac{d}{du}\left( \frac{1}{ D_{\rm 5w(T)} -\frac{m_0}{a} } \right)
{\scriptstyle (-T+a_5;-T+a_5)} P_R
\frac{1}{ D_{\rm 5w(\infty)} -\frac{m_0}{a} } 
{\scriptstyle (-T;-T)} 
\right. \nonumber\\
&&
\left. \quad 
+
P_R \frac{d}{du}\left( \frac{1}{D_{\rm 5w(T)} -\frac{m_0}{a} } \right)
{\scriptstyle (T;T)} \, P_L
\frac{1}{ D_{\rm 5w(\infty)}  -\frac{m_0}{a} } 
{\scriptstyle (T+a_5;T+a_5)} 
\right\} . \nonumber\\
\end{eqnarray}

\subsubsection{Surface term in the limit $T\rightarrow \infty$}

We have seen that 
${\frac{d}{du}\, \overline{\eta}}_{\rm surf}$ is actually
localized at the boundaries $t=-T+a_5$ $(-T)$ and $t=T$ $(T+a_5)$. 
Still it depends on the whole interpolating five-dimensional gauge fields.
We next show that 
in the limit $T\rightarrow \infty$,
the interpolating five-dimensional gauge field 
in the surface contributions
can be replaced by the gauge fields of its boundary values.

In order to show this, let us first introduce 
the five-dimensional gauge fields which are uniform 
with respect to the fifth-dimensional coordinate $t$
\begin{equation}
U^{\leftarrow}_\mu(x,t;u) = U^0_\mu(x), \qquad
U^{\rightarrow}_\mu(x,t;u) = U_\mu(x;u),
\end{equation}
\begin{figure}[htbp]
  \begin{center}

{\unitlength 1mm
\begin{picture}(160,40)
\put(20,0){\line(1,0){85}}
\put(60,-1.5){\line(0,1){3}}
\put(59.5,-5){0}

\multiput(20,0)(5,0){18}{\circle*{1}}
%

{\linethickness{0.5mm}
\put(20,25){\line(1,0){85}}
\put(20,5){\line(1,0){85}} 
\qbezier(50,5)(55,5)(60,15)
\qbezier(60,15)(65,25)(70,25)
}
\multiput(0,5)(5,0){4}{\line(1,0){3}}
\multiput(105,25)(5,0){4}{\line(1,0){3}}

\put(22,8){$U^0_\mu(x)$}
\put(94,28){$U_\mu(x)$}

\put(20,0){\line(0,1){30}}
\put(10,-5){$-T+a_5$}
\put(105,0){\line(0,1){30}}
\put(105,-5){$T$}

\put(42,28){$U^{\rightarrow}_\mu(x,t;u)$}
\put(78,8){$U^{\leftarrow}_\mu(x,t;u)$}
\put(65,15){$U_\mu(x,t;u)$}

\end{picture}
}

    \caption{Uniform five-dimensional gauge fields}
    \label{fig:uniform-5dim-gauge-fields}
  \end{center}
\end{figure}
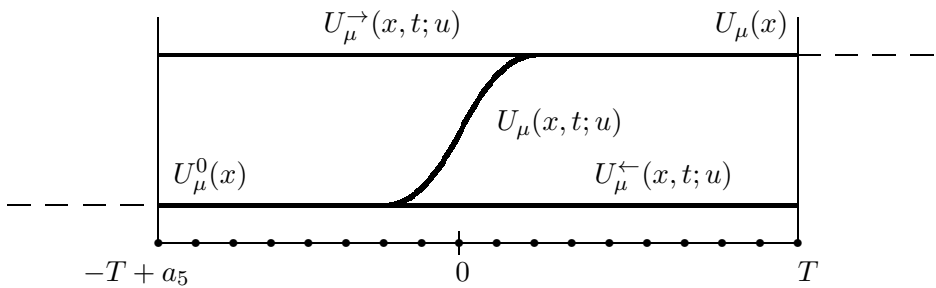

\noindent
and consider the five-dimensional Dirac fermions (the simplified 
domain-wall fermion) which couple to these uniform gauge fields.
We denote the five-dimensional Dirac operator of these fermions 
as $D_{5}^{\leftarrow}$ and $D_{5}^{\rightarrow}$, respectively.

In the contribution from the boundary
at $t=T$ in Eq.~(\ref{eq:eta-variation-surf}), 
the Dirac operator $D_{\rm 5w(T)}$  differs 
from $D_{\rm 5w(T)}^{\rightarrow}$ 
only in the region $t \, a_5 \le  +\Delta$.
Then we may write as
\begin{eqnarray}
&& 
\frac{1}{ D_{\rm 5w(T)}^{\rightarrow} -\frac{m_0}{a} }(T,T) 
- 
\frac{1}{ D_{\rm 5w(T)} -\frac{m_0}{a} } (T,T)
\nonumber\\
&& \quad = 
\sum_{t' \le + \Delta} \frac{1}{ D_{\rm 5w(T)}-\frac{m_0}{a} }(T,t')
\left( D_{\rm 5w(T)} - D_{\rm 5w(T)}^{\leftarrow}\right) (t')
\frac{1}{ D_{\rm 5w(T)}^{\rightarrow}  -\frac{m_0}{a} } (t',T). \nonumber\\
\end{eqnarray}
Since 
both $\left(D_{\rm 5w(T)}-\frac{m_0}{a}\right)^{-1}$ 
and $\left(D_{\rm 5w(T)}^{\rightarrow}-\frac{m_0}{a}\right)^{-1}$ 
decay exponentially at large distance in the fifth dimension,
as shown in Eq.~(\ref{eq:exponetial-bound-on-5dim-D-Dirichlet}),
we can see that the above difference vanishes exponentially 
in the limit $T\rightarrow \infty$.
As for $\left(D_{\rm 5w(\infty)}-\frac{m_0}{a}\right)^{-1}(T+a_5,T+a_5)$, 
the similar results follow from Eq.~(\ref{eq:exponetial-bound-on-5dim-D})
and we may replace it by 
$\left(D_{5(T)}^{\rightarrow}-\frac{m_0}{a}\right)^{-1}(T+a_5,T+a_5)$. 
By the similar argument, we can show that 
$D_{\rm 5w(T)}$ and $D_{\rm 5w(\infty)}$ in
the contribution from the boundary at $t=-T$
can be replaced by $D_{\rm 5w(T)}^{\leftarrow}$ and  
$D_{\rm 5w(\infty)}^{\leftarrow}$, respectively.

Furthermore, since the five-dimensional Wilson-Dirac operators depend
smoothly on the gauge fields, the differences of $D_{\rm 5w(T)}$
vanishes even after taking the variation with respect to the parameter $u$.
Since $U^{\leftarrow}_\mu(x,t;u)$ actually does not depend on $u$, 
this implies that the surface term from the boundary at $t=-T+a_5$
vanishes identically.

Thus we have
\begin{eqnarray}
\label{eq:eta-variation-surf-step1}
&& {\frac{d}{du}\, \overline{\eta}}_{\rm surf} = \nonumber\\
&&
\lim_{a_5 \rightarrow 0} \lim_{T \rightarrow \infty} 
\frac{1}{\pi} {\rm Im} \, \frac{1}{a_5^2} \,
{\rm Tr}_x \, 
P_R \frac{d}{du}
\left( \frac{1}{ D^{\rightarrow}_{\rm 5w(T)} -\frac{m_0}{a} } \right)
{\scriptstyle (T;T)} \, P_L
\frac{1}{ D^{\rightarrow}_{\rm 5w(\infty)}  -\frac{m_0}{a} } 
{\scriptstyle (T+a_5;T+a_5)} .
\nonumber\\
\end{eqnarray}

\subsubsection{
Inverse five-dimensional Wilson-Dirac operators 
in four-dimensional surfaces at boundaries}

We next evaluate the inverse five-dimensional Wilson-Dirac operators
along four-dimensional surfaces, 
which appear in the r.h.s. of Eq.~(\ref{eq:eta-variation-surf-step1}).
The inverse of $\left( D_{5(T)}^\rightarrow-\frac{m_0}{a_5} \right)$ 
at $s=t=T$ 
is nothing but the propagator of the boundary variables
of the simplified domain-wall fermion, which we discussed in
section~\ref{sec:domain-wall-fermion}.
In the limit $T\rightarrow \infty$,
it can be given in terms of the inverse 
of the effective Dirac operator as follows:
\begin{equation}
\label{eq:surface-propagator-at-boundary}
\lim_{T\rightarrow \infty} 
P_R 
\frac{1}{a_5} \,
\frac{1}{ D_{\rm 5w(T)}^\rightarrow -\frac{m_0}{a} } 
{\scriptstyle(T;T)}
= P_R \left( \frac{1}{a D_{\rm eff}} - 1 \right)
\left[ U_\mu(x;u)\right] .
\end{equation}
The inverse of $\left(D_{5(\infty)}^\rightarrow-\frac{m_0}{a_5} \right)$ 
at $s=t=T+a_5$ can be also related to the effective Dirac operator
through its representation 
in terms of the inverse five-dimensional Wilson-Dirac operator,
Eq.~(\ref{eq:effective-Dirac-operator-Intro}). We have
\begin{equation}
\label{eq:surface-propagator}
\frac{1}{a_5}
\frac{1}{ D_{5(\infty)}^\rightarrow -\frac{m_0}{a} } 
{\scriptstyle (T+1;T+1)} P_R
= \frac{a}{2} 
\left( \gamma_5 D_{\rm eff} \gamma_5 - D_{\rm eff} \right)
\left[ U_\mu(x;u)\right]    P_R .
\end{equation}

With these results, the surface term can be evaluated further as 
\begin{eqnarray}
\label{eq:eta-variation-surf-step2}
{\frac{d}{du}\, \overline{\eta}}_{\rm surf}
&=&
\lim_{a_5 \rightarrow 0} 
\frac{1}{\pi} {\rm Im} \,
{\rm Tr}_x \, 
P_R \frac{d}{du}
\left( \frac{1}{ D_{\rm eff}} - 1 \right) 
\, P_L
\frac{1}{2} 
\left( \gamma_5 D_{\rm eff} \gamma_5 - 
D_{\rm eff} \right) \nonumber\\
&=&
\lim_{a_5 \rightarrow 0} 
\frac{1}{\pi} {\rm Im} \,
{\rm Tr}_x \, 
\frac{d}{du} \frac{1}{D_{\rm eff}} \, P_L \, 
\left(- D_{\rm eff} \right)
\ \left[ U_\mu(x;u)\right] \nonumber\\
&=&
\lim_{a_5 \rightarrow 0} 
\frac{1}{\pi} {\rm Im} \,
{\rm Tr}_x \, 
P_L\, \frac{d}{du} D_{\rm eff} \ \frac{1}{D_{\rm eff}} \ 
\left[ U_\mu(x;u)\right]  .
\end{eqnarray}
In the limit $a_5 \rightarrow 0$, $D_{\rm eff}$ reduces to 
the original Neuberger's Dirac operator 
$D$ of Eq.~(\ref{eq:overlap-dirac-operator}) and finally we 
obtain Eq.~(\ref{eq:variation-eta-result}).

\subsubsection{Bulk term in the continuum limit}

We next turn to the bulk contribution 
${\frac{d}{du} \, \overline{\eta}}_{\rm bulk}$ given by
Eq.~(\ref{eq:eta-variation-bulk}),
\begin{eqnarray}
{\frac{d}{du} \, \overline{\eta}}_{\rm bulk}
&=& \lim_{a_5 \rightarrow 0}\lim_{T\rightarrow \infty} 
\frac{1}{\pi} {\rm Im} {\rm Tr}_{(T)} \, 
\frac{d}{du} D_{\rm 5w} \frac{1}{ D_{\rm 5w(\infty)} -\frac{m_0}{a} } .
\nonumber
\end{eqnarray}
We will calculate this contribution in the classical continuum 
limit $a\rightarrow 0$ and will show that 
it reproduces the variation of the Chern-Simons term
up to the local current of Bardeen and Zumino \cite{bardeen-zumino}.

In evaluating the bulk contribution,
we set $a_5=a$. At the same time as to take the continuum limit,
we also take the limit of the infinite extent 
of the fifth dimension $N\rightarrow \infty$,
keeping $T=N a_5 \gg \Delta$ finite. 
The limit $T \rightarrow \infty$ is taken at last.
We adopt the plane wave basis $e^{i k_M (x_M/a) }$, where
the five-dimensional coordinate is denoted as
$x_M=(x_\mu,t)$ $(M=1,2,3,4,5)$ with upper case Latin indices.

The five-dimensional Wilson-Dirac operator $D_{5(\infty)}$ acts 
on the plane wave basis as follows:
\begin{eqnarray}
&& \left( 
D_{5(\infty)}-\frac{m_0}{a}
\right)   e^{i k_M  (x_M/a) }  \nonumber\\
&=&
e^{i k_M (x_M/a) } 
\left( 
\sum_{M=1}^5
\frac{1}{a}
\left( i \gamma_M \sin k_M + 2 \sin^2 \frac{k_M}{2} \right) 
-\frac{m_0}{a} 
\right. \nonumber\\
&& \left. \qquad \qquad \qquad 
     -\sum_M \frac{1}{2} 
\left[(1-\gamma_M)e^{i k_M }  \nabla_M 
     +(1+\gamma_M)e^{-i k_M } \nabla_M^\dagger  \right]
\right) . \nonumber\\
\end{eqnarray}
The second term in the r.h.s. may be expanded for a smooth background as 
\begin{eqnarray}
V(k) &\equiv& \sum_M \frac{1}{2}
\left[(1-\gamma_M)e^{i k_M}  \nabla_M 
                  +(1+\gamma_M)e^{-i k_M} \nabla_M^\dagger \right]
\nonumber\\
& = &
i \frac{\partial}{\partial k_M} S(k)^{-1} \mathcal{D}_M +
\mathcal{O}(a) ,
\end{eqnarray}
where $\mathcal{D}_M$ is the covariant derivative in the continuum limit
\begin{equation}
\mathcal{D}_M = \partial_M+ i A_M(x) 
\end{equation}
and $S(k)$ is the free propagator of the five-dimensional 
Wilson-Dirac fermion,
\begin{equation}
\label{eq:wilson-dirac-propagator}
S(k)^{-1}
=\sum_{M=1}^5
\left( i \gamma_M \sin k_M + 2 \sin^2 \frac{k_M}{2} \right) 
-m_0
\quad (0 < m_0 < 2) .
\end{equation}

Then, inserting the delta-function
\begin{equation}
\delta_{xy}= \int^{\pi}_{-\pi} 
\frac{d^5 k}{(2\pi)^5} \, 
e^{i k_M (x-y)_M/a }
\end{equation}
into the functional trace of the bulk term, 
we obtain the expansion
\begin{eqnarray}
\label{eq:eta-variation-bulk-step1}
{\frac{d}{du} \, \overline{\eta}}_{\rm bulk}
&=& 
\lim_{T'\rightarrow \infty} \lim_{a\rightarrow 0}
\frac{1}{\pi} {\rm Im} 
{\rm Tr}_{(T)} \,  \int^\pi_{-\pi} \frac{d^5 k}{(2\pi)^5} \, \times
\nonumber\\
&& \qquad\qquad
e^{-i k_M (x_M/a) }
\frac{d}{du} D_{5} \frac{1}{ D_{5(\infty)} -\frac{m_0}{a} }
e^{i k_M (x_M/a) } \nonumber\\
&=& 
\lim_{T'\rightarrow \infty} \lim_{a\rightarrow 0}
\frac{1}{\pi} {\rm Im} \, 
\sum_{x_M} \, 
\int^\pi_{-\pi} \frac{d^5 k}{(2\pi)^5} \, \times
\nonumber\\
&& \qquad \qquad 
{\rm Tr} \, (-1) 
\frac{d}{du} V(k) \,
\sum_{l=0}^\infty a^{l+1} \left\{ S(k) V(k) \right\}^l  S(k)
\nonumber\\
&=& 
\lim_{T' \rightarrow \infty} 
\frac{1}{\pi} {\rm Im} \, 
\int \int^{T'}_{-T'} d^4 x dt \ C_{JMNKL}
{\rm Tr} \left\{
\frac{d}{du} A_J \, 
\mathcal{D}_M \mathcal{D}_N \mathcal{D}_K \mathcal{D}_L
\right\} \nonumber\\
&& + \mathcal{O}(a) ,  
\end{eqnarray}
where
\begin{equation}
C_{JMNKL}= 
\int^\pi_{-\pi} \frac{d^5 k}{(2\pi)^5} \, 
{\rm tr} (S \partial_J S^{-1}) 
(S \partial_M S^{-1}) (S \partial_N S^{-1}) (S \partial_K S^{-1}) 
(S \partial_L S^{-1})(k) .
\end{equation}
This coefficient can be evaluated using the fact that 
it gives a topological number associated with
the five-dimensional Wilson-Dirac propagator $S(k)$.
\cite{5dim-CS-in-domain-wall-fermion,kikukawa-yamada}.
It is completely anti-symmetric and takes the following value:
\begin{equation}
C_{JMNKL}= \epsilon_{JMNKL} \frac{i}{8(\pi)^2} ,
\end{equation}
with the convention for the gamma matrices 
$\gamma_1\gamma_2\gamma_3\gamma_4\gamma_5= 1$.
Then the bulk term in the classical continuum limit is given as
\begin{eqnarray}
\label{eq:eta-variation-bulk-step2}
{\frac{d}{du} \, \overline{\eta}}_{\rm bulk}
&=& 
\lim_{T' \rightarrow \infty} 
- \frac{1}{\pi} \,
\int d^4 x \int^{T'}_{-T'} dt \  \frac{1}{32(\pi)^2}
{\rm Tr} \left\{ \frac{d}{du} A_J \, F_{MN} F_{KL} \right\} 
+ \mathcal{O}(a).   \nonumber\\
\end{eqnarray}

\section{The lattice $\eta$-invariant and the effective action for chiral 
Ginsparg-Wilson fermions}
\label{sec:relation-to-effective-action}
\reseteqnum

\subsection{Relation to the effective action for chiral 
Ginsparg-Wilson fermions}

Now we discuss the relation of our lattice implementation of the 
$\eta$-invariant to the effective action for the chiral
Ginsparg-Wilson fermions in abelian and non-abelian chiral
gauge theories \cite{overlap,abelian-chiral-gauge-theory,suzuki}. 
For the one-parameter family of the gauge fields of 
Eq.~(\ref{eq:one-family-4dim-gauge-fields}), 
the effective action for the right-handed chiral Ginsparg-Wilson
fermion is parametrized by $u$:
\begin{equation}
  \overline{\Gamma}_{\rm eff}
= \ln \det M_{kj} \left[ U_\mu(x;u) \right]  .
\end{equation}
The variation of the effective action with respect to the parameter $u$
is obtained from Eq.~(\ref{eq:effective-action-variation}) 
by the choice of 
$\zeta_\mu(x)= \frac{d}{du} U_\mu(x;u) \, U_\mu^{-1}(x;u)$ as 
\begin{equation}
\label{eq:effective-action-variation-u}
  \frac{d}{du} \overline{\Gamma}_{\rm eff}\left[ U_\mu(x;u) \right]
=  {\rm Tr} \frac{d}{du} D \, \hat P_R D^{-1} P_L 
+ \sum_k \left( v_k, \frac{d}{du} v_k \right) .
\end{equation}
Comparing this with the variation of the lattice $\eta$-invariant 
Eq.~(\ref{eq:variation-eta-result}), 
we obtain
\begin{eqnarray}
\frac{d}{du} \overline{\Gamma}_{\rm eff}\left[ U_\mu(x;u) \right]
&=& \frac{d}{du}\, \overline{\eta}\left[U_\mu(x,t;u)\right]
\nonumber\\
&& 
-\lim_{a_5 \rightarrow 0} \lim_{T \rightarrow \infty} 
\frac{1}{\pi} {\rm Im} {\rm Tr}_{(T)} \, 
\, \frac{d}{du} D_{5(\infty)} \, \frac{1}{ D_{5(\infty)} -\frac{m_0}{a}}
\nonumber\\
&& 
\quad + \sum_k \left( v_k, \frac{d}{du} v_k \right) .
\end{eqnarray}
This equation implies that the combination of the last two terms
of the r.h.s. can be writen as a total derivative in $u$ of 
a certain functional of the five-dimensional gauge field, $U_\mu(x,t;u)$.
We denote it by $ 2\pi \overline{Q}_5 \left[U_\mu(x,t;u)\right]$.
Then, by integrating in $u$, 
we obtain a formula for the imaginary part of the effective action:
\begin{equation}
\label{eq:eta-invariant-and-effective-action-lattice}
{\rm Im} \overline{\Gamma}_{\rm eff}\left[U_\mu\right] - 
{\rm Im} \overline{\Gamma}_{\rm eff}\left[U^0_\mu\right]  
= \pi \overline{\eta}\left[U_\mu(x,t)\right]  
+ 2\pi \overline{Q}_5 \left[U_\mu(x,t)\right] ,
\end{equation}
where
\begin{eqnarray}
\label{eq:chern-simons-term-lattice}
2\pi \overline{Q}_5 \left[U_\mu(x,t)\right] 
&\equiv& 
-\lim_{a_5 \rightarrow 0} \lim_{T \rightarrow \infty} 
\int^1_0 du \, 
 {\rm Im} {\rm Tr}_{(T)} \, 
\, \frac{d}{du} D_{\rm 5w(\infty)} 
\, \frac{1}{ D_{\rm 5w(\infty)} -\frac{m_0}{a}}
\nonumber\\
&& \quad
+ \int^1_0 du \, \sum_k \left( v_k, \frac{d}{du} v_k \right) .
\end{eqnarray}
This is the relation which may be regarded as the lattice counterpart 
of Eq.~(\ref{eq:eta-invariant-and-effective-action-continuum}).

$\overline{Q}_5$ here can be regarded as a lattice expression 
of the Chern-Simons term in the following sense. 
1) First of all, 
$\overline{Q}_5$ compensates the dependence 
of $\overline{\eta}$ on the path of the interpolation 
and make it integrable so that
it can give the effective action of chiral fermions,
which depends on only the values of gauge fields at the boundaries.
2) $\overline{Q}_5$ reproduces the non-abelian gauge 
anomaly of the effective action, 
while $\overline{\eta}$ is gauge invariant.
If $U_\mu(x)$ is obtained from $U^0_\mu(x)$  by a certain
gauge tranformation,
\begin{equation}
U_\mu(x) = g(x) U^0_\mu(x) g(x+\hat \mu a)^{-1},
\end{equation}
we may consider an interpolation of the gauge transformation
function, $g(x,t)$, such that 
$ g(x,t=-\infty)=1 $ and $ g(x,t=\infty)=g(x)$ 
and the region of the interpolation is within 
$t\in[-\Delta,\Delta]$.\footnote{In this equation, 
the fifth component of the interpolating five-dimensional
gauge field is introduced. The evaluation 
of $\frac{d}{du}\overline{\eta}$  in section~\ref{sec:variation} 
holds true even if we introduce the fifth component of the gauge 
field as long as its support is within the region 
$t\in[-\Delta, \Delta]$.}
Then we obtain
\begin{eqnarray}
\label{eq:wess-zumino-term}
&&  {\rm Im} \, \overline{\Gamma}_{\rm eff} 
\left[g(x) U^0_\mu(x) g(x+\hat \mu a )^{-1} \right]
-{\rm Im} \, \overline{\Gamma}_{\rm eff} \left[U^0_\mu(x)\right]
\nonumber\\
&& \qquad \quad
= 
2\pi \overline{Q}_5 
\left[g(x,t) U^0_\mu(x) g(x+\hat \mu a,t)^{-1}, 
      g(x,t) g(x,t+a_5)^{-1} \right] . \nonumber\\
\end{eqnarray}

We should also note the role of the contribution of 
the second term of the r.h.s. of Eq.~(\ref{eq:chern-simons-term-lattice}),
so called the measure term.
As we have seen, by virtue of the measure term, 
the path-dependece in the $u$-integration is removed and
$\overline{Q}_5$ becomes a functional of $U_\mu(x,t)$.
It corresponds to the local current of Bardeen and 
Zumino in the continuum theory.
More importantly, 
as shown by L\"uscher in \cite{abelian-chiral-gauge-theory},
the measure term plays a crucial role for 
the gauge invariance of the effective action in abelian
chiral gauge theories on the lattice.

\subsection{Gauge invariance of the lattice Chern-Simons 
term in abelian chiral gauge theories}

The gauge-invariant choice of the measure term
implies the gauge-invariance of 
the lattice Chern-Simons term, $\overline{Q}_5$.
In order to see this, let us consider 
the lattice Chern-Simons term, $\overline{Q}_5$, in 
an abelian gauge theory and examine its 
gauge transformation property 
under the infinitesimal gauge transformation,
\begin{equation}
  \delta A_\mu(x,t) = - \partial_\mu \omega(x,t), 
\end{equation}
\begin{equation}
  \omega(x,t=-\infty)=0, \qquad \omega(x,t=\infty)=\omega(x) .
\end{equation}
Since $\overline{\eta}$ is gauge invariant, 
the transformation of the five-dimensional bulk term of $\overline{Q}_5$ 
(See Eq.~(\ref{eq:chern-simons-term-lattice}))
can be evaluated through the transformation of the surface contribution
in Eq.~(\ref{eq:variation-eta-result}).
Then $\overline{Q}_5$ is transformed as follows:
\begin{eqnarray}
&& \delta_\omega 2\pi \overline{Q}_5
\left[U_\mu(x,t),U_5(x,t)=1\right] \nonumber\\
&&= - \delta_\omega \, \int^1_0 du \, {\rm Im} \,  
{\rm Tr}_x \, P_L\, \frac{d}{du} D \, \frac{1}{D} 
-i \int^1_0 du \, 
a^4 \sum_x \omega(x) 
\partial_\mu^\ast j_\mu(x) \left[ u A_\mu \right]  \nonumber\\
&&= - i\int^1_0 du 
\sum_x \omega(x) \left\{ {\rm tr} \gamma_5\left(1-aD\right)(x,x)
                         - a^4 \partial_\mu^\ast j_\mu(x)
                       \right\}\left[ u A_\mu \right] . 
\end{eqnarray}
Here we have noted that 
$\delta_\omega D\left[ u A_\mu \right]= i u \left[ \omega, D \right]$.

As discussed in section~\ref{sec:effective-action}, 
for anomaly free abelian chiral theories, 
the anomalous term which is induced from the five-dimensional bulk
term can be written in the form
\begin{equation}
{\rm tr} \gamma_5\left(1-aD\right)(x,x) = a^4 \partial_\mu^\ast 
\bar k_\mu(x).  
\end{equation}
On the other hand, it is possible to choose the 
measure term so that it satisfies the anomalous conservation law
of Eq.~(\ref{eq:anomalous-conservation}), 
\begin{equation}
\partial_\mu^\ast j_\mu(x) = \partial^\ast_\mu \bar k_\mu(x).
\end{equation}
Then these two terms cancels exactly 
and the lattice Chern-Simons term $\overline{Q}_5$ becomes
gauge-invariant.

As pointed out by Suzuki in \cite{suzuki}, the ansatz for the 
measure term Eq.~(\ref{eq:ansatz-measure-term})
can be obtained from the following integral expression for 
the effective action:
\begin{equation}
\label{eq:integral-for-effective-action-lattice}
\overline{\Gamma}_{\rm eff} \left[ A_\mu \right] 
= 
\int^1_0 dt \left( 
{\rm Tr} \, P_L\, \frac{d}{dt} D \, 
\frac{1}{D}\left[ t A_\mu \right] 
-i a^4 \sum_x
A_\mu(x) \bar k_\mu(x)
\left[ t A_\mu \right] 
\right).
\end{equation}
Comparing Eq.~(\ref{eq:integral-for-effective-action-lattice})
with Eq.~(\ref{eq:effective-action-variation-u}), we find 
that the integration of the measure term can be given 
by the integration of the local current $\bar k_\mu(x)$:
\begin{equation}
 \int^1_0 du \, \sum_k \left( v_k, \frac{d}{du} v_k \right) 
=- \int^1_0 du \, 
a^4 \sum_x
 A_\mu(x) \bar k_\mu(x)\left[ u A_\mu \right] .
\end{equation}
Then we obtain a compact expression for the lattice Chern-Simons term 
as follows:
\begin{eqnarray}
\label{eq:chern-simons-term-lattice-abelian}
2\pi \overline{Q}_5 \left[U_\mu(x,t)\right] 
&\equiv& 
-\lim_{a_5 \rightarrow 0} \lim_{T \rightarrow \infty} 
\int^1_0 du \, 
 {\rm Im} {\rm Tr}_{(T)} \, 
\, \frac{d}{du} D_{\rm 5w(\infty)} 
\, \frac{1}{ D_{\rm 5w(\infty)} -\frac{m_0}{a}}
\nonumber\\
&& \quad
- \int^1_0 du \, 
a^4 \sum_x
A_\mu(x) \, \bar k_\mu(x)\left[ u A_\mu \right] .
\end{eqnarray}

For the non-abelian gauge theories, the gauge covariant local current 
such like $\bar k_\mu(x)$ in the case of abelian chiral gauge theories 
is not obtained so 
far.\footnote{In the recent work \cite{non-abelian-gauge-anomaly} by 
L\"uscher, it has been shown 
to all orders of an expansion in powers of the lattice spacing 
that the gauge covariant local current of the desired property 
exists. 
}
Such a gauge covariant local current could correct 
the five-dimensional bulk term
\begin{eqnarray}
\label{eq:bulk-term}
\lim_{a_5 \rightarrow 0} \lim_{T \rightarrow \infty} 
\int^1_0 du \, {\rm Im} {\rm Tr}_{(T)} \, 
\, \frac{d}{du} D_{\rm 5w(\infty)} 
\, \frac{1}{ D_{\rm 5w(\infty)} -\frac{m_0}{a}}
\end{eqnarray}
to give the lattice Chern-Simons term $2\pi \overline{Q}_5$ with
desired properties. 

\subsection{Integrability of $\overline{\eta}$ }

In the classical continuum limit, the Chern-Simons term vanishes 
identically, when the condition for gauge anomaly cancellation is 
satisfied. Then the effective action for chiral fermions can be 
given entirely by the $\eta$-invariant. 
Then one may ask whether
this ideal situation would happen 
on the lattice with a finite lattice spacing,
when the condition for gauge anomaly cancellation is satisfied,
\begin{equation}
{\rm Im} \, \Gamma_{\rm eff} \,  \ce^? \, 
\pi {\rm \overline{\eta}}  
\qquad {\rm if} \quad
\sum_R {\rm Tr} \left(T^a \left\{ T^b , T^c \right\}\right) = 0  .
\end{equation}
For this, the five-dimensional bulk term should vanish identically, 
or should become integrable and depend only on the boundary values 
of the interpolating five-dimensional gauge field. It does not seem 
to be the case in general, however, from the result in the previous 
subsection. We need further study on this point.
It may be possible to realize this ideal case by 
deforming the five-dimensional Wilson-Dirac operator, which 
enters to the fermion action 
Eq.~(\ref{eq:five-dimensional-massless-Dirac-fermion}), 
as suggested by Neuberger \cite{geometrical-aspect,math-aspect-overlap}.

\section{Summary and Discussion}
\label{sec:summary}
\reseteqnum

In this paper, we considered a lattice implementation of 
the $\eta$-invariant, using the complex phase 
of the determinant of the (simplified) domain-wall fermion,
which couples to an interpolating five-dimensional gauge field. 
It is realizing the idea of Kaplan and Schmaltz 
explicitly on the lattice. The lattice $\eta$-invariant is examined 
and is shown to have a direct relation to the imaginary part of the 
(gauge invariant) effective action for the chiral Ginsparg-Wilson 
fermion in the case using Neuberger's Dirac operator. 
Although the formula of the lattice $\eta$-invariant seems to be 
practical, the issue of the integrability is remained.
A lattice expression for the five-dimensional Chern-Simons term 
is obtained. 
It should also be examined how the global 
anomaly \cite{global-anomaly} fits in this implementation of 
the $\eta$-invariant \cite{global-anomaly-in-overlap}.

Our analysis shows clearly and explicitly that 
the interplay between the four-dimensional chiral fermion
and the five-dimensional (massless) fermion, which is 
known in the continuum theory, can be realized on the lattice
in the framework of the domain-wall fermion and the overlap
formalism, where the Ginsparg-Wilson relation is built in.
It is expected that other known relations over various dimensions 
could be also realized in the framework of lattice gauge theory.

Quite recently, starting from the Ginsparg-Wilson relation, 
a general formula of the effective action for chiral 
Ginsparg-Wilson fermions is derived 
by L\"uscher \cite{non-abelian-gauge-anomaly}
and its relation to the $\eta$-invariant is suggested. 
It is conceivable that there is a close relation between this 
formula and the implementation of the $\eta$-invariant discussed 
in this paper. The relation should be clarified in detail. 
This issue is under investigation.

\section*{Acknowledgments}
The authors would like to thank H.~Neuberger and H.~Suzuki 
for enlightening discussions. 
Y.K. is also grateful to M.~L\"uscher and
D.B.~Kaplan for discussions and suggestions. 
Y.K. would like to thank T.-W.~Chiu for the kind hospitality 
at Chiral '99 in Taipei. 
Y.K. is supported in part by Grant-in-Aid 
for Scientific Research from Ministry of Education, Science 
and Culture(\#10740116).

\appendix
\section*{Appendix}

\section{Dirichlet boundary condition by surface interaction}
\reseteqnum
\label{app:5d-Dirac-operator-inverse-diff}

In section~\ref{sec:variation}, in order to show
that the surface term $\frac{d}{du} \overline{\eta}$ 
is localized at the boundary, 
we use the following identity which holds for $s,t \in [-T+1,T]$:
\begin{eqnarray}
\label{eq:5d-Dirac-operator-inverse-diff-appendix}
&&
 \frac{1}{ D_{\rm 5w(T)} -\frac{m_0}{a} }\,(s,x;t,y)
-\frac{1}{ D_{\rm 5w(T+\Delta T)} -\frac{m_0}{a} }\,(s,x;t,y)  \nonumber\\
&& \qquad\qquad \qquad\qquad 
=\frac{1}{ D_{\rm 5w(T)} -\frac{m_0}{a} } \, 
V_{(-T+a_5;T)}  \,
 \frac{1}{ D_{\rm 5w(T+\Delta T)} -\frac{m_0}{a} } \, (s,x;t,y) \nonumber\\
\end{eqnarray}
where 
\begin{eqnarray}
\label{eq:surface-interaction-appendix}
V_{(-T+a_5;T)} &=&   \frac{1}{a_5} \left\{
- P_L \delta_{s,-T}\delta_{t,-T+1} 
- P_R \delta_{s,-T+1}\delta_{_t,-T} 
\right.
\nonumber\\
&& 
\left. \qquad \qquad
- P_L \delta_{s,T}\delta_{t,T+1} 
- P_R \delta_{s,T+1}\delta_{t,T} \right\}.
\end{eqnarray}
In this appendix, we give the derivation of this identity.

For this purpose, let us introduce the five-dimensional Dirac 
fermion defined in the larger five dimensional space
$[-T-\Delta T+a_5,T+\Delta T]$, but with the couplings between
the lattice sites $(-T, -T+a_5)$ and between the lattice sites 
$(T, T+a_5)$ omitted.

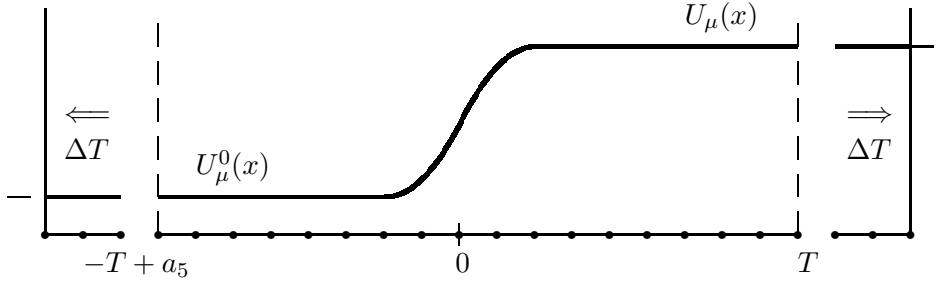
\begin{figure}[htbp]
  \begin{center}
{\unitlength 1mm
\begin{picture}(160,40)

\put(5,0){\line(1,0){10}}
\put(20,0){\line(1,0){85}}
\put(110,0){\line(1,0){10}}
\put(60,-1.5){\line(0,1){3}}
\put(59.5,-5){0}

\multiput(5,0)(5,0){24}{\circle*{1}}
\put(5,0){\line(0,1){30}}
\multiput(20,0)(0,5){6}{\line(0,1){3}}
\put(120,0){\line(0,1){30}}
\multiput(105,0)(0,5){6}{\line(0,1){3}}
\put(10,-5){$-T+a_5$}
\put(105,-5){$T$}
\put(7.5,15){$\Longleftarrow$}
\put(7.5,10){$\Delta T$}
\put(111.5,15){$\Longrightarrow$}
\put(111.5,10){$\Delta T$}

{\linethickness{0.5mm}
\put(5,5){\line(1,0){10}}
\put(20,5){\line(1,0){30}}
\put(70,25){\line(1,0){35}} 
\put(110,25){\line(1,0){10}} 
\qbezier(50,5)(55,5)(60,15)
\qbezier(60,15)(65,25)(70,25)
}
\multiput(0,5)(5,0){3}{\line(1,0){3}}
\multiput(110,25)(5,0){3}{\line(1,0){3}}

\put(25,8){$U^0_\mu(x)$}
\put(90,28){$U_\mu(x)$}

\end{picture}
}
    \caption{Implementation of Dirichlet B.C. by surface interaction}
    \label{fig:dirichlet-bc-by-surface-interaction}
  \end{center}
\end{figure}

\noindent
We denote the five-dimensional Dirac operator of this system by 
$D^\vee_{5(T+\Delta T)}$. Then the difference between
$D^\vee_{5(T+\Delta T)}$ and $D_{5(T+\Delta T)}$ is given by 
the surface interaction $V_{(-T+a_5;T)}$ of 
Eq.~(\ref{eq:surface-interaction}) 
(Eq.~(\ref{eq:surface-interaction-appendix})).
\begin{equation}
D^\vee_{\rm 5w(T+\Delta T)} 
= D_{\rm 5w(T+\Delta T)}  - V_{(-T+a_5;T)} .
\end{equation}

Then we have
\begin{eqnarray}
&& \frac{1}{D^\vee_{\rm 5w(T+\Delta T)}-\frac{m_0}{a}} 
-\frac{1}{D_{\rm 5w(T+\Delta T)}-\frac{m_0}{a}}  \nonumber\\
&& \qquad\qquad\qquad\qquad\qquad\qquad
=\frac{1}{D^\vee_{\rm 5w(T+\Delta T)}-\frac{m_0}{a}} \, V_{(-T+1;T)} \, 
\frac{1}{D_{\rm 5w(T+\Delta T)}-\frac{m_0}{a}} . \nonumber\\
\end{eqnarray}
On the other hand, the field variables in the interval $[-T+a_5,T]$ does not
have any coupling to those outside the region and they are nothing but
the field variables described by $D_{\rm 5w(T)}$. Then,
we have 
\begin{equation}
  \frac{1}{D^\vee_{\rm 5w(T+\Delta T)} -\frac{m_0}{a}} \,(s,x;t,y)
= \frac{1}{ D_{\rm 5w(T)} -\frac{m_0}{a}} \,(s,x;t,y)   
\end{equation}
for $s,t \in [-T+a_5,T]$.
From these two relations,
Eq.~(\ref{eq:5d-Dirac-operator-inverse-diff-3})
(Eq.~(\ref{eq:5d-Dirac-operator-inverse-diff-appendix}))
follows immediately.
In the limit $\Delta T \rightarrow \infty$, it reduces to 
Eq.~(\ref{eq:5d-Dirac-operator-inverse-diff}).

\end{document}